\documentclass[11pt,letterpaper]{article}
\pdfoutput=1
\usepackage{jheppub}

\usepackage{amsmath}
\usepackage{amssymb}
\usepackage{xcolor}
\usepackage{graphicx}
\usepackage{subfig}
\usepackage{bbold}
\usepackage{tabularx}
\usepackage[makeroom]{cancel}
\usepackage{multirow}
\usepackage{slashed}
\usepackage{hyperref}
\usepackage{comment}

\usepackage{tikz}
\usetikzlibrary{calc,fadings,decorations.pathreplacing}
\usepackage{verbatim}

\usepackage{graphicx}
\usepackage{hyperref}

\newcommand{\pd}{\partial}

\title{Chern-Simons bubbles: Lopsided false vacuum decay in axion electrodynamics}

\author{Saquib Hassan,}
\author{John March-Russell,}
\author{and Georges Obied}

\emailAdd{saquib.hassan@physics.ox.ac.uk} 
\emailAdd{john.march-russell@physics.ox.ac.uk}
\emailAdd{georges.obied@physics.ox.ac.uk}

\affiliation{Rudolf Peierls Centre for Theoretical Physics \\University of Oxford \\OX1 3PU Parks Road \\ Oxford \\ United Kingdom}

\date{\today}

\definecolor{darkgreen}{RGB}{50,150,0}
\newcommand{\go}[1]{\textcolor{darkgreen}{GO: #1}}

\definecolor{darkred}{RGB}{150,0,50}
\newcommand{\sh}[1]{\textcolor{darkred}{SH: #1}}

\definecolor{darkblue}{RGB}{0,50,150}

\abstract{
We study axion electrodynamics, including the Chern-Simons interaction term, in the presence of parallel background electric and magnetic fields, as can for example occur in certain models of axion inflation and in the study of dyonic black holes. In this setup, we find a new back-reacted instanton solution which corresponds to the nucleation of an axion domain wall that screens the electromagnetic fields in a process analogous to Schwinger pair production, despite the absence of light charged particles. The full solution includes the effect of the Chern-Simons induced charges and currents on the axion domain wall arising from the Witten and Sikivie (anomalous Hall) effects, respectively.
The Euclidean solution has a reduced $O(2)\times O(2)$ symmetry which describes the nucleation of a prolate bubble in its rest-frame. A unique feature of this solution is that the region of lower energy density is outside the bubble rather than inside. We also describe the time evolution of this initial configuration, showing how the bubble can become further elongated along the direction of the background electric and magnetic fields. We describe potential applications of this process in particle physics and cosmology. 
}

\begin{document}

\maketitle


\section{Introduction}
\label{sec:Intro}

Axion electrodynamics, the theory of an Abelian gauge field interacting with a neutral pseudo-scalar axion via an anomaly term, is a theory rich with both phenomenological implications and theoretical interest. In particular, the presence of a topological Chern-Simons interaction between the axion and the Abelian gauge field (``photon" from now on\footnote{Though here we are mainly interested in theories that do not explicitly include massive charged matter, so this Abelian gauge field is not precisely the Standard Model photon.}) leads to a variety of striking features. For example, this term leads to monopoles gaining electric charge in an asymptotically-constant background axionic field \cite{Witten:1979ey,Fischler:1983sc}.  Moreover, axion domain walls can carry electric charges and currents, thereby generating magnetic fields as well \cite{Sikivie:1984yz,Sikivie:1984bp,Wilczek:1987mv}, and, in the $(3+1)d$ case with $(2+1)d$ world-volume domain walls the currents are now understood as a manifestation of the (fractional) quantum Hall effect (FQHE) states that generally exist on the domain walls (see for example~\cite{Hidaka:2021kkf}). 

These features are of special interest to us in this work where we study a class of non-perturbative effects in axion electrodynamics. These effects facilitate the decay of background electromagnetic fields due to a novel form of the Schwinger process involving not the charged particle or monopole states, which are extremely heavy and not present in the effective theory, but, rather, the nucleation of axion domain wall bubbles.  Due to the Chern-Simons interaction term, the domain wall displays the Witten and Sikivie (anomalous Hall) effects, picking up charge and current densities in the presence of the background electric and magnetic fields (see section~\ref{sec:WittenSikivie}).  This allows the decay of metastable states of the combined $a,{\bf E},{\bf B}$ system, simultaneously changing the axion vacuum expectation value (vev) and screening the background electromagnetic field. As such, we find that background parallel electric and magnetic fields induce nucleation of domain wall bubbles that interpolate from one value of the axion field to another where the vacuum energy is lower. Although this is superficially similar to simple cases of false vacuum decay, we find many aspects of the dynamics to be quite unusual. Most notably, due to the Chern-Simons term and associated Witten and Sikivie effects and the long-range nature of the $U(1)$ interaction
the region of lower energy density is \emph{outside} the axion domain wall, and the  post-nucleation real time evolution is highly, and increasingly asymmetric.

More concretely, our focus will be on axion electrodynamics with an action:
\begin{equation}\label{eq:AEDaction}
    I=\int d^4 x \left( -\frac{1}{4}F_{\mu \nu}F^{\mu \nu} - \frac{f^2}{2}\partial_{\mu}\theta \partial^{\mu}\theta - V(\theta) +  \frac{K\alpha}{8\pi} \theta \epsilon_{\mu \nu \rho \sigma} F^{\mu \nu}F^{\rho \sigma} \right),
\end{equation}
where $F_{\mu\nu}=\pd_\mu A_\nu - \pd_\nu A_\mu$ is the Maxwell field strength corresponding to the photon $A_\mu$, $\theta \equiv a/f$ where $a$ is the pseudo-scalar axion field with periodicity $a\sim a+2\pi f$, $f$ is the mass-dimension one axion decay constant and $\alpha=e^2/4\pi$ is the $U(1)$ fine structure constant. The integer $K$ specifies the Chern-Simons interaction term between the gauge field and the axion. Note that $\epsilon_{\mu \nu \rho \sigma}$ is the totally antisymmetric symbol with $\epsilon_{0123}=+1$.

Since we are mainly interested in the physically most motivated case of a massive axion we also include a non-zero axion potential, $V(\theta)$. Our conclusions will apply to any $V(\theta)$ that is a periodic function.
This includes the case where $V(\theta)$ has a sub-periodicity $V_N(\theta) = V_N(\theta + 2\pi/N)$ corresponding to a $\mathbb{Z}_N$ global symmetry
of the theory with $N$ degenerate local minima.\footnote{When UV completed to a theory including gravity we expect that this $Z_N$ global symmetry to be explicitly broken. The size of this symmetry breaking can be exponentially smaller than the non-perturbative effects that give rise to $V_N(\theta)$, and we can work in the approximation where this breaking is sub-leading to the effects of the background gauge fields.}  In phenomenological studies of the QCD axion, the integer $N \geq 1$ is known as the domain wall number and is typically determined by anomalies of the UV theory.  For concreteness we will consider a simple potential of the ``one-instanton" form (although we emphasize that our conclusions apply to more general periodic potentials):
\begin{equation}
\label{eq:axionpotential}
 V(\theta) \equiv V_N(\theta) = \left(\frac{mf}{N}\right)^2 \Big(1-\cos(N \theta)\Big),
\end{equation}
where we have chosen the normalization of $V_N(\theta)$ so that the parameter $m$ is the mass of the axion around any one of the minima of $V_N(\theta)$.  The theory is thus specified by the two integers $K,N$, the dimensionless coupling $\alpha$ and the dimensional parameters $m,f$.

The action eq.~\eqref{eq:AEDaction} can be thought of as arising from a UV theory of the axion interacting with the photon in the presence of heavy $U(1)$-charged fermions which transform chirally under the continuous global shift symmetry $a(x)\rightarrow a(x) +c$ for $c\in \mathbb{R}$ a constant.  These fermions determine the integer $K$ in the Chern-Simons topological coupling through their anomaly coefficient. In addition the axion is coupled to a sector with non-perturbative dynamics that generates a potential for the axion breaking the continuous shift symmetry $a(x)\rightarrow a(x) + c$ down to a $\mathbb{Z}_N$ subgroup. See also~\cite{DiLuzio:2020wdo} for a review of these effects. We will not need to specify the exact form of these dynamics as long as it takes place at energy scales much higher than $m$, in other words, well above the low-energy effective field theory regime containing the axion and the photon degrees of freedom.

We are mostly interested in the case where there are parallel background electric and magnetic fields. In this situation, $\epsilon_{\mu \nu \rho \sigma} F^{\mu \nu}F^{\rho \sigma} = 8\mathbf{E\cdot B}\neq 0$, which leads to a monodromy (see for example~\cite{McAllister:2014mpa}) breaking of the discrete shift symmetry in $\theta$. Then, moving between adjacent minima can change the energy in the electromagnetic fields and there are transitions from one vacuum to another of lower energy. We will compute the nucleation rate of such processes both in the thin wall and thick wall limits.

Among other things, such nucleation processes can play an important role in early universe physics where they can lead to a screening of electromagnetic fields. An example is in the inflationary model of Anber-Sorbo~\cite{Anber:2009ua} where the action is exactly of the form given in equation~\eqref{eq:AEDaction}. In this model, the inflaton rolls down a steep potential but is slowed down by converting its kinetic energy to electromagnetic radiation through the CS coupling and this radiation is in turn diluted by the inflationary expansion. The conversion works by producing a non-zero $\langle \mathbf{E\cdot B} \rangle$ that is coherent on large (sub-horizon) scales. Towards the end of inflation, these electromagnetic fields can play a role in preheating through a phenomenon known as gauge preheating~\cite{Adshead:2015pva}. However, the presence of charged matter can alter this process due to screening of these gauge fields by Schwinger pair production~\cite{Cado:2022pxk}. Similarly, it is plausible that this background also allows for the nucleation of axion domain walls that also screen the electromagnetic fields. This has the feature that it requires no further assumption on the presence of light charged particles since the axion is already part of the theory~\eqref{eq:AEDaction}.

Of course, the ``electromagnetic" fields in~\eqref{eq:AEDaction} could be in a hidden sector in which case, the domain wall nucleation process might be relevant even later in the history of the universe due to a different source of $\mathbf{E\cdot B}$ fields (see for instance~\cite{Berghaus:2020ekh} for scenarios where dark radiation has significantly different behaviour from that of the visible sector).

In fact, the axion domain wall nucleation we describe occurs more generally than in the theory shown in~\eqref{eq:AEDaction}. The two most straightforward generalizations are to non-Abelian and higher form gauge fields (e.g. the 4-form gauge field used in~\cite{Kaloper:2008fb}) and we claim there is an analogous domain wall nucleation process in these theories as well. This is easy to see by considering the following weak coupling argument which will be made more precise in section~\ref{sec:O4solution}. In all these theories, the equations of motion of the gauge fields reduce to the free equations in the limit where the CS coupling is taken to vanish. As such, these equations can be solved by constant gauge field strengths. On the other hand, the axion equation does not necessarily decouple in this limit because we can simultaneously make the field strengths large to keep the coefficient of the $\theta$ term in the potential constant. In such a limit, the gauge fields provide a `rigid' contribution that tilts the axion potential due to the linear CS coupling. This allows the axion to tunnel from one minimum to another reducing its energy. Intuitively, this process will also happen without taking the strict zero-coupling large-field limit. In these cases, there could also be applications to related models of inflation such as chromo-natural inflation that relies on a background non-Abelian field strength~\cite{Adshead:2012kp} and axion monodromy which has a CS coupling to a 4-form~\cite{Kaloper:2008fb}. 

In addition, the decay rate we calculate will also apply to branes that couple to gauge fields in the same way as an axion domain wall. It is important to understand the ways in which these can contribute to the dynamics of the early universe. In a companion paper, we will study this brane nucleation process in Nariai black hole backgrounds and argue for a bound on the tension of such branes in de Sitter space.

Finally, and as we will see in more detail in sections~\ref{sec:O2O2solution} and~\ref{sec:timeEvolution}, our bubble nucleation and evolution does not preserve the maximal rotational symmetry of the vacuum. This is expected since the presence of parallel electric and magnetic fields already breaks this symmetry. In fact, as we show in section~\ref{sec:O2O2solution}, back-reaction effects cause the nucleated domain walls to be elongated along the preferred direction set by the parallel electric and magnetic fields. The bubble expansion then enhances this asymmetry (see section~\ref{sec:timeEvolution}). This is a distinct feature of the process we are studying. In particular, one could contrast this behavior to that studied in~\cite{Adams:1989su} where they describe how perturbations with low wave number freeze out as a conventional bubble expands. (This can be seen from studying the de Sitter world-volume theory of the expanding bubbles~\cite{Garriga:1991ts,Garriga:1991tb}. It would be interesting to see how the large asymmetry we observe in our simulations can be explained in terms of a world-volume theory on the expanding domain wall.)
As a consequence of this asymmetry in the
bubble growth, it is in principle possible to have an additional source of gravitational wave production  beyond what is usually considered in first order phase transitions, and, moreover, a possibility of primordial black hole formation in the collision of just two such bubbles. These possibilities deserve a dedicated study which we intend to return to in a later work. We briefly comment on some aspects of the asymmetric growth in the conclusions. 

In this work, we initiate these various lines of study by describing the domain wall nucleation process in the theory of~\eqref{eq:AEDaction}. This paper is organized as follows. In Section \ref{sec:4dBubbles} we turn to the main novel results of our study where we show that in the presence of background fields, axion domain wall nucleation can occur. First, neglecting back-reaction, we find a leading order solution that is $O(4)$ symmetric in Euclidean space-time coordinates in section~\ref{sec:O4solution}. Then, we include back-reaction effects that break this symmetry group to $O(2)\times O(2)$ and discuss this solution in section~\ref{sec:O2O2solution}. We find this solution numerically and describe how it screens the electromagnetic fields. Afterwards, in section~\ref{sec:timeEvolution}, we show the initial bubble configuration and its non-linear time evolution including corrections to the fields due to induced charges and currents on the bubble itself. This is where the effects described in \cite{Witten:1979ey,Fischler:1983sc,Sikivie:1984yz,Wilczek:1987mv} play a crucial role. Finally we conclude in section~\ref{sec:Conclusions} with implications of these findings, as well as new applications of these results that are in preparation. Appendix~\ref{sec:2dBubbles} discusses the physics of the non-perturbative process in $(1+1)d$ electrodynamics in the presence of a background electric field, which shares some features of the $(3+1)d$ case. This serves to provide an analogy with the Schwinger process~\cite{Heisenberg:1936nmg,Schwinger:1951nm} in a simple setting. 
The following appendix~\ref{sec:back-reaction} elaborates on back-reaction effects in a weak coupling expansion.

\section{Chern-Simons bubble nucleation in $(3+1)d$}
\label{sec:4dBubbles}
In this section, we will discuss the main result of our paper which is the nucleation and evolution of axion domain walls. We begin by reviewing the Witten and Sikivie effects in section~\ref{sec:WittenSikivie}. These are the effects that will be responsible for screening the electromagnetic fields. Then in section~\ref{sec:O4solution} we describe domain wall nucleation without taking into account this screening. This will be done by taking an appropriate `no back-reaction' limit. This analysis gives the correct leading order expression for the decay rate (up to an overall $\mathcal{O}(1)$ coefficient in the thick wall case that we comment on). However, in order to pin down the symmetries and time evolution of the domain wall, we have to include back-reaction due to the Witten and Sikivie effects. We do this in section~\ref{sec:O2O2solution} for the nucleation and then describe the time-evolution in section~\ref{sec:timeEvolution}.

\subsection{Witten and Sikivie effects}
\label{sec:WittenSikivie}

We start by quoting the equations of motion for the axion and photon fields which are obtained by varying the action of eq.~\eqref{eq:AEDaction}, with $V_N(\theta)$ as given in eq.~\eqref{eq:axionpotential}. These equations are
\begin{align}
    &-\partial_{t}^{2}\theta+\partial_{i}^{2} \theta = m^2\left(\frac{1}{N}\sin(N\theta)-K\frac{\alpha}{8\pi}  \left(\frac{1}{mf}\right)^2 \epsilon^{\mu \nu \rho \sigma}F_{\mu \nu}F_{\rho \sigma}\right), \label{eq:eomaxion} \\
    &\partial_{\mu}F^{\nu \mu}=-\frac{K\alpha}{2\pi}\epsilon^{\mu \nu \rho \sigma}\partial_{\mu}(\theta F_{\rho \sigma}), \label{eq:eomphoton}\\
    &\epsilon^{\mu \nu \rho \sigma} \partial_\nu F_{\rho \sigma}=0, \label{eq:Bianchi}
\end{align}
where the last equation is the Bianchi identity. For constant electromagnetic and $\theta$ fields, eq.~\eqref{eq:eomphoton} and eq.~\eqref{eq:Bianchi} are automatically satisfied. If so, we can also satisfy eq.~\eqref{eq:eomaxion} by ensuring that its right hand side vanishes (when the term involving the electromagnetic fields has magnitude smaller than $1/N$). Such values of $\theta$ label the minima and maxima of an ``effective'' potential for constant electric and magnetic field backgrounds. The extrema of this potential are given by solutions to the equation: 
\begin{equation}
\label{eq:potminima}
    \sin(N\theta)= \frac{N K\alpha}{\pi} \frac{\mathbf{E\cdot B}}{(mf)^2} .
\end{equation}
In particular, in a setup with constant $\mathbf{E}$ and $\mathbf{B}$ fields oriented along the $+z$-axis the $F \tilde{F}$ term is non-zero and the axion field is displaced from the origin, also acquiring a non-zero value. In this background, with the axion in a local minimum of its potential, we are interested in discussing a process that corresponds to the tunneling of the axion field under the potential barrier~\cite{Coleman:1977py}. As usual, the tunneling event results in the creation of a bubble inside which the axion is different from (and in our case larger than) its asymptotic value in the local minimum. The boundary of this bubble is an axion domain wall that interpolates between the value of the axion inside the bubble and its value in the local minimum. 
This domain wall bubble is nucleated at rest so all time derivatives vanish initially\footnote{The fields are nevertheless evolved non-trivially in Euclidean time, where a bubble emerges from the vacuum at negative Euclidean time, and upon reaching critical radius at Euclidean time $\tau=0$, then proceeds to expand in real time. Section \ref{sec:O2O2solution} automatically incorporates this point.}. However, as the axion field varies in space, it induces effective current and charge densities in the Maxwell equations due to the CS term on the right hand side of equation~\eqref{eq:eomphoton}. These effects will modify the electromagnetic fields inside and around the bubble. We may compute corrections to the electric and magnetic fields by integrating eq.~\eqref{eq:eomphoton} subject to the Bianchi identity of eq.~\eqref{eq:Bianchi}. In vector component form, eq.~\eqref{eq:eomphoton} reads
\begin{equation}
\label{eq:eomphotonvector}
    \nabla \cdot \mathbf{E}=-\frac{K\alpha}{\pi}\nabla \cdot\left(\theta \mathbf{B}\right), \quad \mathrm{and}\quad  \nabla \times \left(\mathbf{B}-\frac{K\alpha}{\pi}\theta \mathbf{E}\right)=\partial_t\left(\mathbf{E}+\frac{K\alpha}{\pi}\theta \mathbf{B}\right),
\end{equation}
and the Bianchi identity remains the same as in Maxwell electrodynamics. Since all time derivatives vanish on the initial time slice that we are focusing on, we can use the Bianchi identities to reduce equations~\eqref{eq:eomphotonvector} to the simple form:
\begin{align}
    \label{eq:eomEfield}
    \nabla \cdot \mathbf{E} = -\frac{K \alpha}{\pi} \nabla \theta \cdot \mathbf{B}, \quad \mathrm{and}  \quad
    \nabla \times \mathbf{B} = \frac{K\alpha}{\pi}\nabla\theta \times \mathbf{E}.
\end{align}

We can view the first equation in eq.~\eqref{eq:eomEfield} as a modified Gauss Law, with $\rho_{E} \sim  \mathbf{B} \cdot \nabla \theta$. The fact that axion gradients parallel to magnetic fields behave like effective electric charges is a manifestation of the Witten effect~\cite{Witten:1979ey}. In our case, this feature implies that the axion domain wall carries electric charge in a transverse background $\bf B$-field. In particular, when the magnetic field is oriented along the $+z$ direction as in our setup, the axion domain wall on the initial time slice acquires an effective electric charge density $\rho_E \propto \cos \psi$ where $\psi$ is the polar angle measured from the $+z$-axis. The sign of the electric charge is positive in the northern hemisphere and negative in the southern hemisphere so as to produce an electric field that (partially) screens the original electric field inside the bubble. This is similar to what happens in Schwinger pair production~\cite{Schwinger:1951nm}. That said, just outside the poles of the bubble, we expect the electric field to be enhanced compared to its original background value. Finally, we note that the induced charge is proportional to the magnetic field rather than the electric field as happens when polarizing a material.

Similarly, the second of eq.~\eqref{eq:eomEfield} is a modified Amp\`{e}re's law where the cross product on the right hand side represents an effective current density. This effect, whereby axion domain walls carry currents in the presence of parallel background electric fields, has been described a while ago by Sikivie~\cite{Sikivie:1984yz,Sikivie:1984bp} and we will refer to this phenomenon as the Sikivie effect or anomalous Hall effect. Since our background electric field is aligned along the $+z$-axis, the cross product means that the effective current density is of the form $\mathbf{J} \propto \sin \psi \mathbf{\hat{\varphi}}$ where $\varphi$ is the azimuthal angle in 3D space and $\mathbf{\hat{\varphi}}$ is the unit vector in the direction of increasing $\varphi$. This Hall current is localised on the domain wall and has a magnitude that peaks on the equator. This current is always in the $\mathbf{\hat{\varphi}}$-direction and leads to an anti-screening (enhancement) of the magnetic field inside the bubble. The same current, however, leads to a screening of the magnetic field just outside the bubble and near the equator. Again, we note that the value of the current is proportional to the electric field value in the background.

The qualitative discussion above shows that, once back-reaction is taken into account, the electromagnetic fields inside and around the bubble are modified from their original background values. It is then easy to see that this would affect the axion bubble solution itself by considering eq.~\eqref{eq:eomaxion} for example. Of course, the equations ~\eqref{eq:eomaxion},~\eqref{eq:eomphoton} and~\eqref{eq:Bianchi} should all be solved simultaneously, and we will do this below, but let us first briefly comment on what we would expect from such a solution. This discussion will be heuristic but helpful in understanding the nature of the solution we get. 

Going back to the initial time slice, the pattern of (anti-)screening of the electric/magnetic fields discussed above leads to the following observations. First note that the effective charge (current) density leads to a uniform electric (magnetic) field inside the bubble. This is easy to see by analogy with elementary problems in electrodynamics: The distribution of the effective charge density is similar to that of a dielectric sphere in a background electric field which is known to give a uniform electric field inside the sphere. Similarly, the current density is like that of a charged spherical shell rotating at constant angular velocity which gives a uniform magnetic field inside the shell. Deep inside the bubble, the electric and magnetic fields are then constant and so is the value of the axion field which is set by eq.~\eqref{eq:potminima}. 

Now imagine moving from the center of the bubble towards the north pole. All fields remain constant until we get close to the bubble wall. We know that the electric field is larger outside the bubble above the north pole. As such, we expect the axion value to be larger above the north pole. Heuristically one can again see this using eq.~\eqref{eq:potminima} although this is only heuristic because we are ignoring all derivatives. Conversely, we can imagine moving from the center of the bubble towards the equator. All fields are again constant but once we get close to the bubble wall, the magnetic field starts to decrease. The same heuristic argument, using eq.~\eqref{eq:potminima}, now tells us that the axion field should be lower around the equator of the bubble. These two effects, i.e. larger axion field values just outside the poles and smaller field values around the equator, means that bubble is elongated along the $+z$-axis (recall that the inside of the bubble has a larger field value than the outside). We now turn to a more quantitative analysis of this bubble nucleation process.
\subsection{Nucleation with the $O(4)$ approximate instanton}
\label{sec:O4solution}
In this section we study axion domain wall nucleation in the limit where we can ignore the back-reaction on electromagnetic fields. This is the first iteration of the solution describing the nucleation process and we will improve this in the following section by including back-reaction effects. Said differently, the Witten and Sikivie effects discussed in the previous section that lead to screening of the electromagnetic fields and elongation of the nucleated bubbles are ignored. As mentioned previously, we will calculate the (thin-wall) bubble nucleation rate and the critical bubble radius in this approximation which is sufficient to get the correct parametric dependence and provide an analytic understanding of the process. These limits will be made more precise below. 


We will use Coleman's Euclidean prescription~\cite{Coleman:1977py} to calculate the domain wall nucleation rate and the critical bubble size. As such, we first Wick rotate the action~\eqref{eq:AEDaction} to Euclidean signature (taking $\tau=it$) to get:
\begin{equation}\label{eq:EuclideanAEDaction}
    I_E=\int d^4 x \left( \frac{1}{4}F_{\mu \nu}F^{\mu \nu} + \frac{f^2}{2}\partial_{\mu}\theta \partial^{\mu}\theta + V_N(\theta) -i  \frac{K\alpha}{8\pi} \theta \epsilon_{\mu \nu \rho \sigma} F^{\mu \nu}F^{\rho \sigma} \right),
\end{equation}
where $\epsilon_{123\tau} = +1$ (note the difference from the Lorentzian signature in eq.~\eqref{eq:AEDaction}). Also, $iA_{\tau}=A_{0}$, so the components of $F_{\mu\nu}$ here are not the same as in Lorentzian signature. With this action, the equations of motion become:
\begin{align}
    &\partial_{\tau}^{2}\theta+\partial_{i}^{2} \theta = m^2\left(\frac{1}{N}\sin(N\theta)- i K\frac{\alpha}{8\pi}  \left(\frac{1}{mf}\right)^2 \epsilon^{\mu \nu \rho \sigma}F_{\mu \nu}F_{\rho \sigma}\right), \label{eq:Euclideaneomaxion} \\
    &\partial_{\mu}F^{\nu \mu}=i\frac{K\alpha}{2\pi}\epsilon^{\mu \nu \rho \sigma}\partial_{\mu}(\theta F_{\rho \sigma}), \label{eq:Euclideaneomphoton}\\
    &\epsilon^{\mu \nu \rho \sigma} \partial_\nu F_{\rho \sigma}=0. \label{eq:EuclideanBianchi}
\end{align}
Once more, these equations are easy to satisfy with constant electromagnetic and $\theta$ fields and these solutions correspond to, when rotated back to Lorentzian signature, the solutions we discussed in section~\ref{sec:WittenSikivie}. In Euclidean signature, the field strength tensor has imaginary electric field components and real magnetic field components.

We will consider the case where the electromagnetic fields are identically constant. More precisely, we take a limit where we can ignore the back-reaction of the axion field on the electromagnetic fields. This can be done by taking $\alpha \rightarrow 0$ and $F_{\mu\nu} \rightarrow \infty$ keeping the product $\alpha \epsilon^{\mu \nu \rho \sigma}F_{\mu \nu}F_{\rho \sigma}$ finite. In this limit, the right hand side of eq.~\eqref{eq:Euclideaneomphoton} vanishes but the term in eq.~\eqref{eq:Euclideaneomaxion} arising from the Chern-Simons interaction remains relevant. We can then study axion electrodynamics in the presence of non-dynamical background electromagnetic fields which we now do. We will show in section~\ref{sec:O2O2solution} that this limit is a consistent leading order solution.

For concreteness, we orient our coordinate system such that the background electric and magnetic fields are $\mathbf{E}=E\mathbf{\hat{z}}$ and $\mathbf{B}=B\mathbf{\hat{z}}$ respectively. Since the parallel background electric and magnetic fields spontaneously break the shift symmetry in $\theta$, the potential when we are ignoring the dynamics of the electromagnetic field is 
\begin{equation} \label{eq:EuclideaneffectiveV}
    \frac{V(\theta)}{m^2 f^2} =\frac{1}{N^2} \Big(1-\cos(N \theta)\Big) - K \frac{\alpha E B}{\pi m^2 f^2} \theta.
\end{equation}
The extrema are given by solutions to the equation: 
\begin{equation}
\label{eq:Euclideanpotminima}
    \sin(N\theta)=N K \frac{\alpha E B}{\pi m^2 f^2}, 
\end{equation}
where we will study the case in which the right hand side has magnitude less than unity so that there are static non-rolling solutions. There is then a transition process in which the false vacuum at one minimum of the potential tunnels to a lower minimum. In this process, a background value of $\theta$ exists in space, and bubbles of higher values of $\theta$ with lower potential are spontaneously nucleated. We now solve for the bubble profile and compute the nucleation rate in the special case of the thin wall limit.

To work in the thin wall regime, we consider parameters where $NK\alpha E B / m^2 f^2 \ll 1$ and $N\theta \ll 1$ (while still working in the limit of negligible back-reaction). Then a minimum exists at $\theta = \theta_0$ which is given by a solution of eq.~\eqref{eq:Euclideanpotminima}
\begin{align}
\label{eq:minimumvalue}
\theta_0 \approx K \frac{\alpha E B}{\pi m^2 f^2},
\end{align}
with a subsequent minimum at $\theta = \theta_0+2\pi/N$. Following~\cite{Coleman:1977th}, we expect that the solution with highest symmetry has the lowest action and therefore dominates the nucleation process. As such, we look for an $O(4)$ symmetric instanton solution for the axion so that $\theta=\theta(u)$ with 
\begin{align}
u=m\sqrt{\tau^2+x^2+y^2+z^2}
\end{align}
the dimensionless radial coordinate. Then eq.~\eqref{eq:eomaxion} becomes approximately
\begin{equation}\label{eq:o4eom}
    \frac{d^2 \delta\theta}{du^2}+\frac{3}{u}\frac{d \delta\theta}{du}=\frac{1}{N}\sin(N\delta\theta)+\mathcal{O}(\theta_0)
\end{equation}
where we have defined $\delta\theta=\theta(u)-\theta_0$ and expanded the right hand side in the limit $\theta_0 \ll 1$ as follows from eq.~\eqref{eq:minimumvalue}. This gives the equation of motion for a particle rolling in a potential with dimensionless time $u$ in the presence of a damping force inversely proportional to $u$. By standard arguments~\cite{Coleman:1977py}, we can ignore the friction term in the thin wall limit and we find that the sine-Gordon soliton is an appropriate solution:
\begin{equation}\label{eq:sineGordan4d}
    \theta(u)=\theta_0+\frac{2\pi}{N}-\frac{4}{N}\arctan{\left(\exp{(u-u_*)}\right)}.
\end{equation}
The wall thickness is therefore $\mathcal{O}(m^{-1})$ and the wall is situated at $x = m^{-1} u_*$. We may find $u_*$ by substituting this solution in~\eqref{eq:EuclideanAEDaction} and minimizing to obtain the critical bubble size:
\begin{align}
\label{eq:4dcriticalradius}
u_* \approx \frac{12 m^2 f^2}{\alpha KN EB}.
\end{align}
Moreover, using the critical radius we can compute the value of the Euclidean action which directly gives the nucleation rate for such a thin-wall bubble:
\begin{equation}
\label{eq:nucleationrate}
     \Gamma/V \sim \exp{\left(- \frac{6912\pi^2 m^4 f^8}{(\alpha K E B)^3  N^5}\right)}.
\end{equation}
This is one of the main results of this paper along with the study of the back-reacted solution in the next section and the real-time evolution in section~\ref{sec:timeEvolution}.

\subsection{Back-reaction and the $O(2)\times O(2)$ instanton}
\label{sec:O2O2solution}
We now turn to the fully back-reacted solution. Due to the directionality of the background fields, and as per the discussion in section~\ref{sec:WittenSikivie}, we should not expect the fully back-reacted instanton to be $O(4)$-symmetric as in the case without back-reaction. The back-reaction we find only preserves an $O(2) \times O(2)$ subgroup of the full $O(4)$: the first factor rotates the $\tau z$-plane, and the second rotates the $xy$-plane. This is the instanton we will look for in this section. As we will see, knowing this symmetry helps inform our coordinate and gauge choices making the problem more tractable. 

Assuming that the back-reacted instanton solution preserves this $O(2) \times O(2)$ symmetry, we define dimensionless radial coordinates in the $xy$- and $\tau z$-planes:
\begin{align}
r &= m\sqrt{x^2+y^2}, \quad
\rho = m\sqrt{\tau^2+z^2}.
\end{align}
In this coordinate system, the fields in Euclidean signature are simply $\theta=\theta(\rho,r)$, and $A=A_\rho(\rho,r)d\rho+A_\sigma(\rho,r)d\sigma+A_r(\rho,r)dr+A_\phi(\rho,r)d\phi$. We use $\sigma$ and $\phi$ to denote the angular coordinates in the $\tau z$- and $xy$- planes respectively and all fields are independent of $(\phi, \sigma)$ by symmetry. We now have the $O(2)\times O(2)$ ansatz. 

It is easier to solve for the gauge fields rather than the field strength directly. The Bianchi identity is now $\epsilon^{\mu\nu\alpha\beta}\nabla_{\nu}F_{\alpha \beta}=0$, and if the gauge field depends only on the radial coordinates, then one can show that the Binachi identity is automatically satisfied. In addition, we have the freedom to pick a convenient gauge choice. Our ansatz above already assumes that the gauge fields are independent of the angular coordinates so that we may no longer perform gauge transformations that depend on the angular coordinates since these spoil the ansatz. That said, we may still consider $A\rightarrow A+d\Lambda (\rho,r)$. We now pick $\Lambda (\rho,r)$ such that $A_r$ vanishes. This gauge choice eliminates one of the field components; we can still perform gauge transformations that depend on $\rho$ only but this may not be enough to eliminate $A_\rho (\rho, r)$. 
Instead, we use an equation of motion, eq.~\eqref{eq:Euclideaneomphoton}:
\begin{equation}\label{eq:nablaF}
    \nabla_\mu F^{\nu \mu}=i\frac{K\alpha}{\pi}\epsilon^{\mu \nu \alpha \beta}\partial_\mu \theta \partial_\alpha A_\beta.
\end{equation}
Upon inspecting the $\nu=\rho$ and $r$ components\footnote{The covariant derivative is important here due to our coordinates. The left hand side of eq.~\eqref{eq:nablaF} can be evaluated using the $\nabla_\mu F^{\nu \mu}=\partial_\mu F^{\nu \mu} +\Gamma^{\mu}_{\mu \sigma}F^{\nu \sigma}+\Gamma^{\nu}_{\mu \sigma}F^{ \sigma \mu} \rightarrow \partial_\mu F^{\nu \mu} +\Gamma^{\mu}_{\mu \sigma}F^{\nu \sigma}$ since $F_{\mu \nu}$ is anti-symmetric. Furthermore, the remaining Christoffel symbol can be expanded to give $\nabla_\mu F^{\nu \mu}=h^{-1/2} \partial_\mu (h^{1/2}F^{\nu \mu})$ where $h$ is the determinant of the metric $h_{\mu \nu}=\mathrm{diag}(1,r^2,1,\rho^2)$ in the $O(2)\times O(2)$ adapted coordinate system.} we find that $A_\rho=k_1 \rho^{-1}\log(k_2 r)$, subject to the gauge choice $A_r=0$, where $k_1$ and $k_2$ are integration constants. Imposing Neumann boundary conditions at $\rho=0$, we deduce that $A_\rho=0$. This is another component of the gauge field that now vanishes. Therefore, we may obtain the entirety of the information regarding the electric and magnetic fields by solving for $\theta, A_\phi,$ and $A_\sigma$. 

We will be calculating in the weak coupling limit in which $\alpha \ll 1$. It is useful to organize the equations in powers of $\alpha$. To this end, we rescale $A_\mu\rightarrow \sqrt{\alpha}A_\mu/m$. The Euclidean equations of motion~\eqref{eq:Euclideaneomaxion} and~\eqref{eq:Euclideaneomphoton} are now: 
\begin{align}
    \partial_\rho^2 \theta+\frac{\partial_\rho \theta}{\rho}+ \partial_r^2 \theta+\frac{\partial_r \theta}{r}-\frac{\sin(N\theta)}{N}+\frac{i K}{\pi r\rho}\left(\frac{m}{f}\right)^2 \left(\partial_\rho A_\sigma \partial_r A_\phi - \partial_r A_\sigma \partial_\rho A_\phi \right)&=0, \label{eq:eomo2o21}\\
    \partial_\rho^2 A_\sigma-\frac{\partial_\rho A_\sigma}{\rho}+ \partial_r^2 A_\sigma+\frac{\partial_r A_\sigma}{r}-\frac{iK\alpha}{\pi}\frac{\rho}{r} \left(\partial_r \theta \partial_\rho A_\phi - \partial_\rho \theta \partial_r A_\phi \right)&=0, \label{eq:eomo2o22}\\
    \partial_\rho^2 A_\phi+\frac{\partial_\rho A_\phi}{\rho}+ \partial_r^2 A_\phi-\frac{\partial_r A_\phi}{r}-\frac{iK\alpha}{\pi}\frac{r}{\rho} \left(\partial_r \theta \partial_\rho A_\sigma - \partial_\rho \theta \partial_r A_\sigma \right)&=0. \label{eq:eomo2o23}
\end{align}
We now have a system of non-linear coupled partial differential equations, and in general we do not expect analytic solutions. The last two equations encapsulate the Witten and Sikivie effects. 
As a sanity check, let us first consider the familiar small back-reaction limit which is the limit $\alpha \rightarrow 0$ in the above equations. We may neglect the non-linear $\mathcal{O}(\alpha)$ terms in equations~\eqref{eq:eomo2o22} and~\eqref{eq:eomo2o23}. Then the background configurations of constant electric and magnetic fields - $E_0$ and $B_0$ respectively - solve these two equations; i.e. 
\begin{align}
\label{eq:EfieldBC}
    i\tilde{A}_\sigma=\frac{\tilde{E}_0 \rho^2}{2m^2}, \quad \mathrm{and} \quad
    \tilde{A}_\phi=\frac{\tilde{B}_0 r^2}{2m^2}.
\end{align} where we have denoted $\tilde{E_0}=\sqrt{\alpha}E_0$ and $\tilde{B_0}=\sqrt{\alpha}B_0$. 
We may then substitute these expressions in eq.~\eqref{eq:eomo2o21}, which reduces to 
\begin{equation}
    \partial_\rho^2 \theta+\frac{\partial_\rho \theta}{\rho}+ \partial_r^2 \theta+\frac{\partial_r \theta}{r}-\frac{\sin(N\theta)}{N} + \frac{K}{\pi}\left(\frac{m}{f}\right)^2 \frac{\tilde{E_0} \tilde{B_0}}{m^4}=0.
\end{equation}
This is readily solved by the $O(4)$ symmetric ansatz \cite{Coleman:1977th}: if one takes $\theta(\rho,r)=\theta(\sqrt{\rho^2+r^2})$, then it follows that $\partial_\rho^2 \theta+\rho^{-1}\partial_\rho \theta+ \partial_r^2 \theta+r^{-1}\partial_r \theta\rightarrow \partial_u^2 \theta +3u^{-1}\partial_u \theta$, where now $u=\sqrt{\rho^2 +r^2}$. We therefore obtain
\begin{equation}\label{0thordersolution}
    \frac{d^2 \theta}{du^2} +\frac{3}{u}\frac{ d \theta}{du}-\frac{\sin(N\theta)}{N} + \frac{K}{\pi}\left(\frac{m}{f}\right)^2 \frac{\tilde{E_0} \tilde{B_0}}{m^4}=0,
\end{equation} 
at $\mathcal{O}(\alpha^0)$ which we may solve numerically, or with a phenomenological approximation to the wall profile, or with the thin wall approximation under appropriate circumstances. We call this $O(4)$ symmetric solution $\tilde{\theta}(u)$. It is by solving this expression that we may obtain the bounce exponent in $\Gamma/V \sim \exp{(-B)}$, such as in section~\ref{sec:O4solution}.\footnote{We note that numerical solutions show that even thick wall cases possess the same bounce exponent as in eq.~\eqref{eq:nucleationrate} up to an overall $\mathcal{O}(1)$ multiplicative factor. Nevertheless, an $\mathcal{O}(1)$ reduction in the exponent can lead to a significant enhancement of the nucleation rate.} Incorporating back-reaction effects should then give $\mathcal{O}(\alpha)$ corrections to the exponent. It is also now clear why the procedure of section~\ref{sec:O4solution} was justified: the regime in which $\alpha\ll 1$, but $\tilde{E_0}$ and $\tilde{B_0}$ are held fixed (eq.~\eqref{eq:EfieldBC}), gives a critical size to the bubble which depends only on $\tilde{E_0}\tilde{B_0}=\alpha E_0 B_0$, as we see in eq.~\eqref{eq:4dcriticalradius}. Back-reaction effects are corrections on top of this $\mathcal{O}(\alpha^0)$ profile. Typically, we can only solve the set of equations~\eqref{eq:eomo2o21},~\eqref{eq:eomo2o22} and~\eqref{eq:eomo2o23} numerically subject to specific boundary conditions, as we now discuss (some details are in appendix \ref{sec:back-reaction}).

\subsection{Example solutions}

We will now, utilizing some toy choice of parameters, demonstrate both thin wall and thick wall critical bubble solutions at nucleation and
explore some of their features. In section~\ref{sec:timeEvolution} we will discuss the subsequent real time evolution of the nucleated bubbles.

\subsubsection{Thin wall}\label{sec:thinwall}

\begin{figure}[h]
\centering\includegraphics[width=0.5\textwidth]{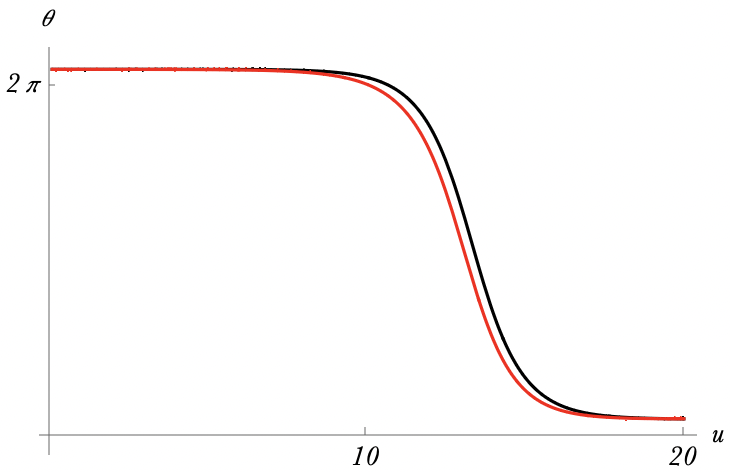}
\caption{Bubble profile for $\tilde{\theta}(u)$ at leading order. The red curve denotes the numerical solution, and the black curve denotes the thin wall approximation (see text).}
\label{fig:0thordersolutionvsthinwall}
\end{figure}

Consider the choice $\alpha=0.1, m/f=0.1$, with anomaly coefficient and domain wall number set $K=N=1$, as well as the background fields $E_0/m^2=B_0/m^2=30$. The asymptotic value of the axion is $\theta_0\approx 0.291$, while its interior value is $\theta_0+2\pi$. For the leading order $O(4)$ symmetric profile, we must solve eq.~\eqref{0thordersolution}. In this regime the thin wall approximation is applicable since using eq.~\eqref{eq:4dcriticalradius} one finds $u_*\approx 13.3 \gg 1$, and as shown in figure~\ref{fig:0thordersolutionvsthinwall}, the numerical solution is very well approximated by the thin wall solution from eq.~\eqref{eq:sineGordan4d}.
\begin{figure}[h]
\centering\includegraphics[width=0.9\textwidth]{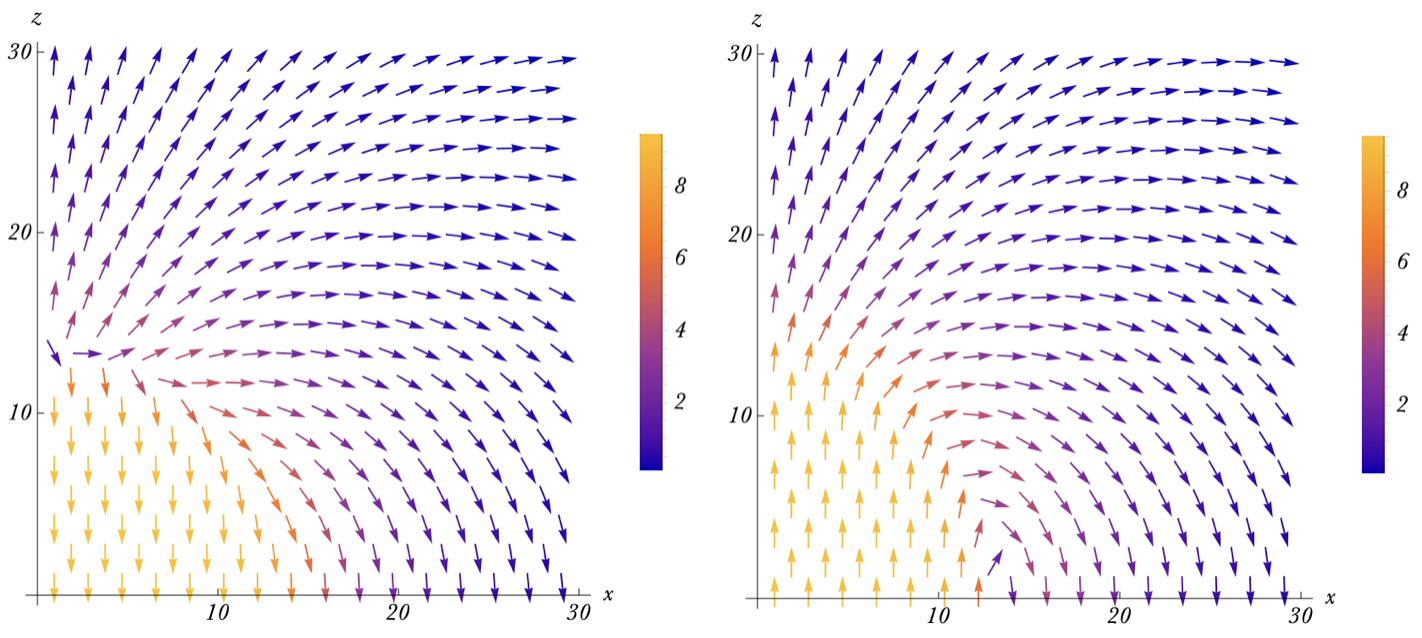}
\caption{The left panel shows the electric field sourced by the nucleated axion domain wall; similarly the right panel shows the magnetic field induced by the domain wall currents (both fields divided by $\alpha$). The constant $\mathbf{\hat{z}}$-directed background $\mathbf{E}$ and $\mathbf{B}$ fields are not shown. The induced fields partially screen the background $\mathbf{E}$ field inside the domain wall and less so
outside around the equator, while anti-screening the $\mathbf{E}$ field outside the poles. The $\mathbf{B}$ fields are anti-screened inside the domain
wall and somewhat less so outside the poles of the domain wall, but screened significantly in a toroidally shaped equatorial region outside the wall.
}
\label{fig:EBfields}
\end{figure}
\begin{figure}[h]
\centering\includegraphics[width=0.9\textwidth]{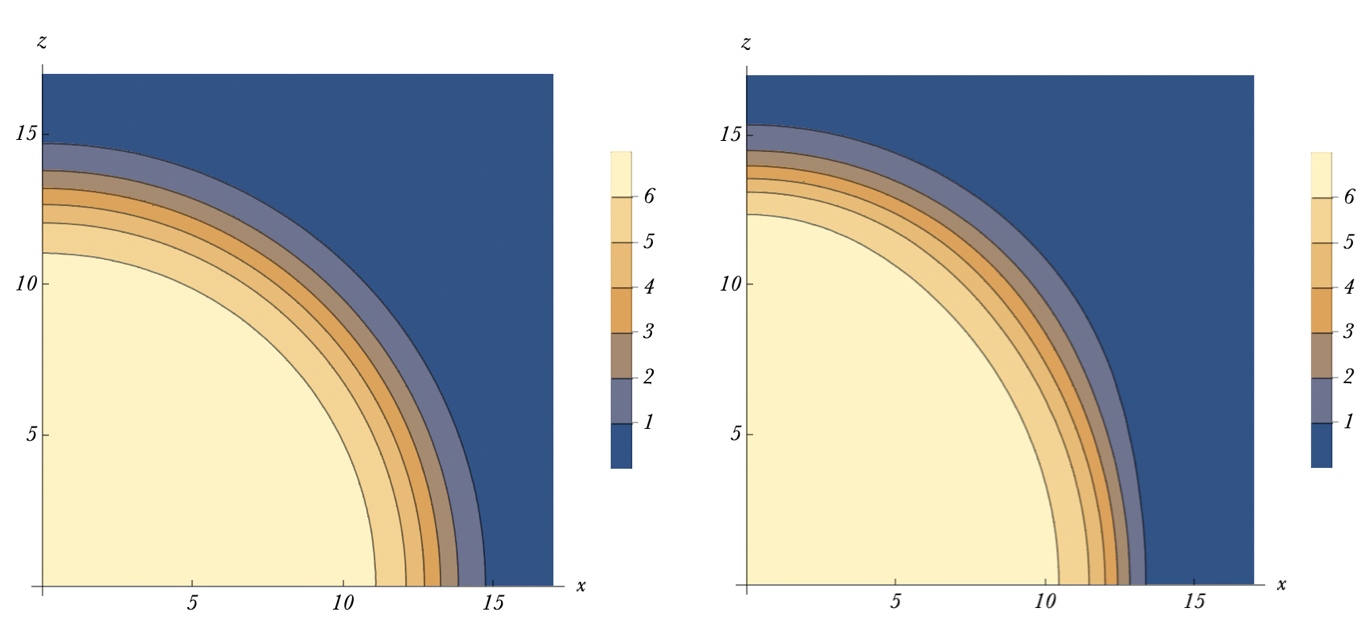}
\caption{Left panel shows the leading-order O(4)-symmetric bubble profile $\tilde{\theta}(u)$, so not including back-reaction effects from the domain-wall-generated electric and magnetic fields. Contour lines indicate the value of the axion field varying from exterior value $\theta_0$ to interior value $\theta_0+2\pi$ (we have taken $N=1$). Right panel shows the bubble profile including leading-order back-reaction effects $\tilde{\theta}(u)+\alpha\delta\theta(\rho,r)$. The bubble has become slightly prolate.
}
\label{fig:bubblecorrection}
\end{figure}
\begin{figure}[h]
\centering\includegraphics[width=0.58\textwidth]{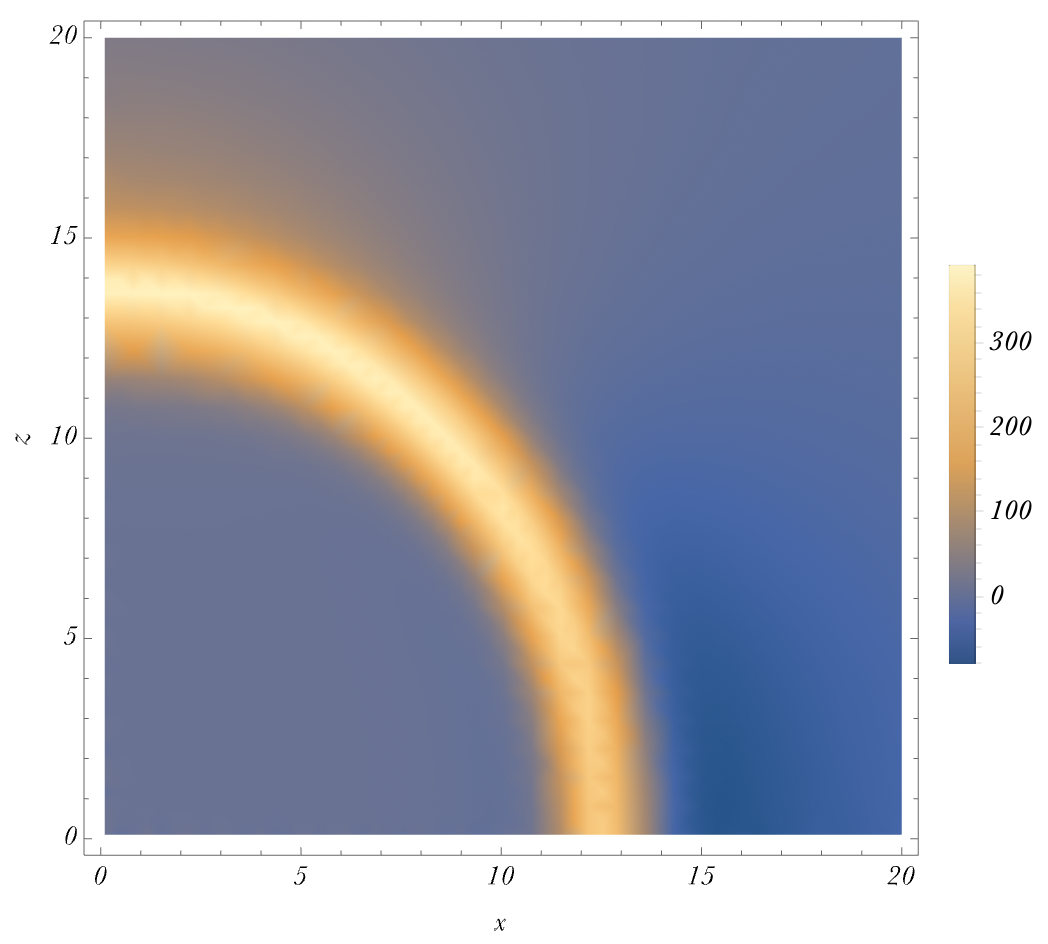}
\caption{Energy density with the uniform background value subtracted. The bright yellow represents the wall. The pale blue in the bubble interior and far in the exterior indicate the same energy density at this order. The region of lower energy density is dark blue, just outside the bubble near the $x$ axis. The pale yellow just above the north pole represents anti-screening (see text). The dark blue region dominates the volume-integrated change in energy upon rotating the figure about the $z$ axis to obtain the full three dimensional picture.}
\label{fig:energydensity}
\end{figure} 
We then use this leading solution in eqs.~\eqref{eq:eomo2o21}-\eqref{eq:eomo2o23} to compute the back-reaction of the electric and magnetic fields, as well as the correction to the axion profile (appendix \ref{sec:back-reaction}). 
Obtaining these solutions for the gauge field components, we then directly compute the electric and magnetic field perturbations. We take the $\tau=0$ slice so $\rho\rightarrow z$, and by rotational symmetry set $y=0$ so that $r\rightarrow x$. See figure \ref{fig:EBfields}. Notice, given the bubble profile, the electromagnetic field pattern is intuitively what we expect based upon the discussion in section \ref{sec:WittenSikivie}. The modified axion profile is shown in figure \ref{fig:bubblecorrection}.

We would also like to observe the energy density distribution in the $xz$ -plane. The wall will carry positive energy due to a varying axion profile of course, but we would like to see where this energy is sourced from. To that end, we use the Hilbert energy momentum tensor:
\begin{align}
\label{eq:EMtensor4d}
    T^{\mu\nu}&=F^{\mu\lambda}F^{\nu}_{\ \lambda} -  f^2 \partial^{\mu}\theta \partial^{\nu}\theta 
    -\eta^{\mu \nu}\left(\frac{1}{4}F_{\alpha\beta}F^{\alpha\beta}+
    \frac{f^2}{2}(\partial\theta)^2 +\left(\frac{m f}{N}\right)^2 (1-\cos{(N\theta)})
    \right),
\end{align}
where we remark that the Chern-Simons term does not explicitly contribute\footnote{The same arguments that we make in appendix \ref{sec:2dBubbles}, including footnotes, carry through here; in particular, this stress tensor agrees with the canonical version despite seeming not to due to Chern-Simons terms that appear in the canonical form. Those terms cancel when simplified, giving agreement with the Hilbert form.}. Evaluating $T^{00}$, we find that the bubble interior has, to leading non-trivial order, \textit{the same energy density as the far exterior}! In contrast, we see that the region of lower energy density is actually \textit{outside} the bubble, like a doughnut - or torus -  about the equatorial plane (see figure \ref{fig:energydensity}). This energy distribution is in stark contrast to the usual version of false vacuum decay in which the interior contains lower energy density, as in \cite{Coleman:1977py} for instance. Moreover, it is different from even the $(1+1)d$ version of appendix \ref{sec:2dBubbles}. As far as we are aware, such a configuration is unlike anything else existing in the literature. 

To explain this energy distribution, note that from eq.~\eqref{eq:EMtensor4d}, the $\theta$-dependent piece changes on the wall and gives positive energy density as anticipated. However, the bizarre energy distribution stems from the electromagnetic part. From figure \ref{fig:EBfields}, above the north pole, both electric and magnetic fields actually enhance the background electric and magnetic fields respectively. In the interior, the electric field is screened and the magnetic field is anti-screened, so the change is energy density cancels at this order, giving the same energy density as the far away false vacuum. Just outside the bubble, about the equatorial plane, figure \ref{fig:EBfields} shows that \textit{both} the electric and magnetic fields are screened, however. It is precisely this zone that has lower energy relative to the background. Despite initial appearances, this region actually has a large volume, which one can picture by rotating the diagrams around the $z$ axis. So there is a torus-shaped region of lower energy density outside the prolate bubble.

\subsubsection{Thick wall}\label{sec:thickwall4d}

Now consider the following choice of parameters: as before, $N=K=1$, but now we set $\alpha=0.1\pi$, $m/f=1/\sqrt{2}$, $E_0/m^2=7$, $B_0/m^2=2$. The background value of the axion field is $\theta_0\approx 0.775$. We find the leading order solution, given by numerically integrating eq.~(\ref{0thordersolution}). The interior value of $\tilde\theta(u)-\theta_0$ as $u\rightarrow0$ is approximately $5.71$. 
\begin{figure}[h]
\centering\includegraphics[width=1.0\textwidth]{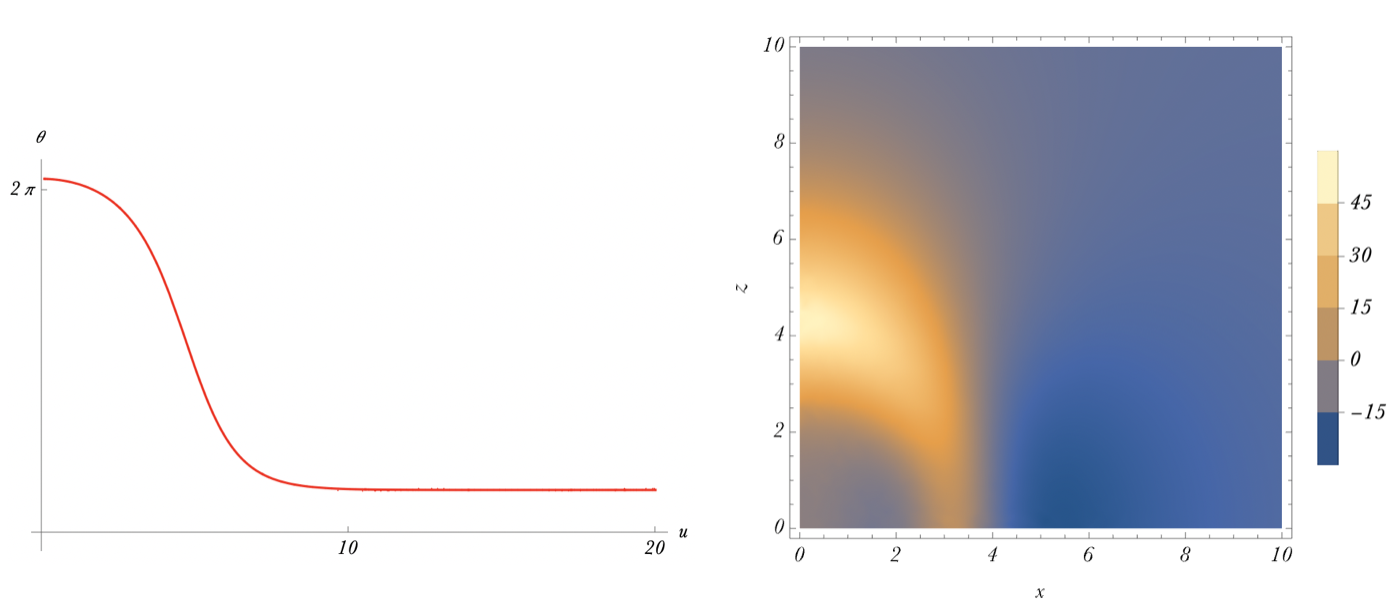}
\caption{The left panel shows the thick wall bubble profile for $\tilde{\theta}(u)$ at leading order and at the instant of nucleation. The right panel shows the full energy distribution, including the back-reacted field profiles computed using $\tilde{\theta}(u)$, at nucleation. Observe again that the lower energy density region is outside the equator of the domain wall in a toroidally shaped region, rather than on the inside.}
\label{fig:thickwallcombined}
\end{figure}
In fact, as the field tunnels, it does not even reach a subsequent minimum, but emerges instead at some intermediate value, from which it will then classically roll in real time. We consider the real time evolution of this bubble in section \ref{sec:timeEvolution}.
Again, using these leading solutions, we can compute the back-reaction effects, which modify the shape of the bubble, using eq.~\eqref{eq:eomo2o21} - eq.~\eqref{eq:eomo2o23}. Incorporating these effects in eq.~\eqref{eq:EMtensor4d} gives the energy density (with the background value subtratcted), shown on the right hand side of figure \ref{fig:thickwallcombined}. The dark blue regions indicate lower energy density.

\section{Real time evolution}
\label{sec:timeEvolution}

The Euclidean evolution that we have been using prepares an initial state on the $t = 0$ time slice in Minkowski space. In section \ref{sec:pressure} we argue qualitatively why the bubble now expands, despite the fact that the lower energy density is now in the bubble exterior. We also look at numerical time evolution using the Lorentzian signature (classical) equations of motion in section \ref{sec:expansion}. In this case, we find that it is numerically more efficient to track the electric and magnetic fields (and the axion) directly since these equations have single time derivatives. The equations we use are the Maxwell equations in~\eqref{eq:eomphotonvector}, the usual electromagnetic Bianchi identities and the axion equation of motion in~\eqref{eq:eomaxion}. 

\subsection{Pressure differences on the bubble wall}\label{sec:pressure}
We can intuitively understand the expansion of the bubble by studying the pressures on either side of the wall. To that end, we use the Hilbert energy momentum tensor eq.~\eqref{eq:EMtensor4d}.
\begin{figure}[h]
\centering\includegraphics[width=1.0\textwidth]{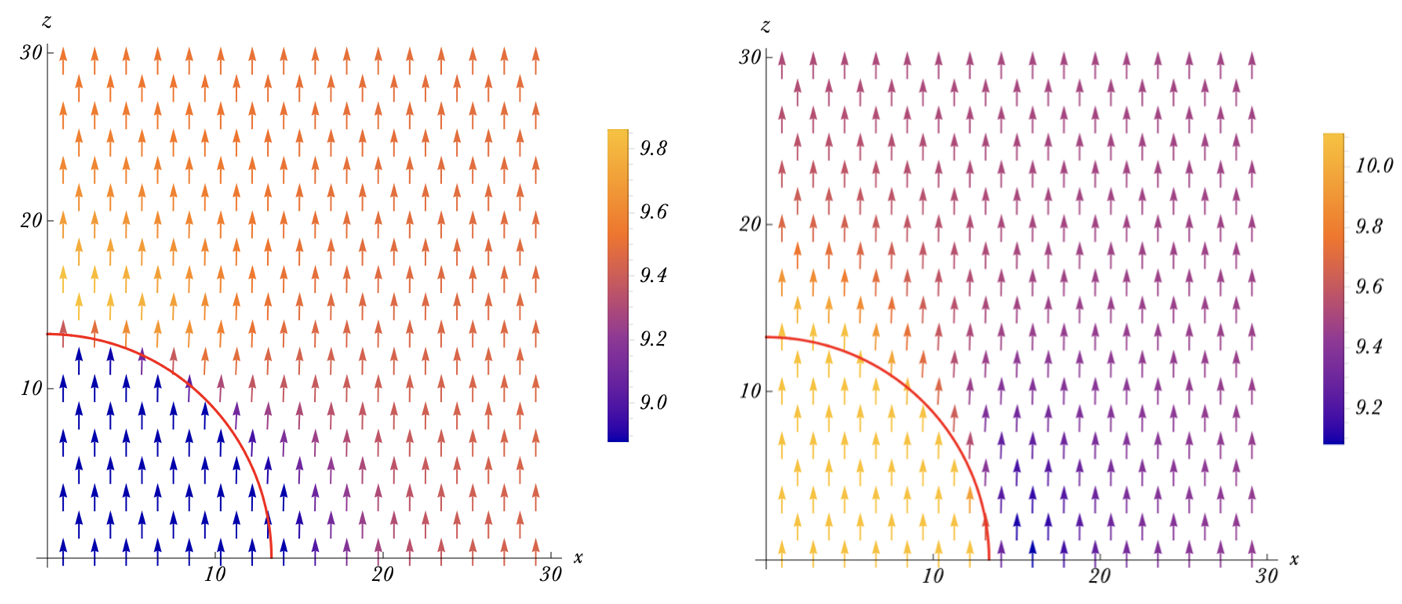}
\caption{The left panel shows the back-reacted electric field from figure \ref{fig:EBfields} superimposed on the background $\tilde{E}_0/m^2$. Similarly, the right panel shows the back-reacted plus the background magnetic field. The red line indicates the wall position (see section \ref{sec:thinwall}).}
\label{fig:EBfields2}
\end{figure}
It is convenient, though not essential, to have the thin wall profile of section \ref{sec:thinwall} in mind (see also figure \ref{fig:EBfields2}). We require the spatial components,
\begin{equation}
    T_{ij}\approx\frac{1}{2}(E^2+B^2)\delta_{ij}-(E_i E_j+B_i B_j).
\end{equation}
Notice that since we are in the thin wall regime, the terms depending on axion derivatives are negligible away from the wall so we do not include them above. Furthermore, to leading non-trivial order in the $\alpha$-expansion, the axion changes by $2\pi$ from one side of the wall to the other (here again we specialize to the $N=1$ case, though similar statements are true for $N>1$ as well), so we can drop the $\cos\theta$ potential term when computing pressure differences across the wall. As such, only the electromagnetic fields are important for this discussion. 

Let us start by comparing the pressures along the $z$-axis on either side of the bubble wall near the north pole, i.e. with $T_{zz}$. Just outside the north pole, the electric field is anti-screened (enhanced) and the magnetic field is also enhanced. On the other side of the domain wall, i.e. just inside the bubble near the north pole, the magnetic field is still anti-screened, but the electric field is screened. A quick computation of $T_{zz}\approx-(E_z^2+B_z^2)/2$ shows that the pressure is higher (i.e. less negative) inside the bubble causing a force in the $+z$-direction (see figure \ref{fig:EBfields2}). So we infer an outward force along the north pole. A similar argument - due to the $O(2)\times O(2)$ symmetry - indicates that there is an outward force along the south pole as well. 
Next, consider the pressure differences near the equator of the bubble in figure~\ref{fig:EBfields2}. For this, consider $T_{rr}$ (in cylindrical coordinates). Just outside the bubble wall, along the $x$-axis or equatorial plane for that matter, $E_x, E_y, B_x,$ and $ B_y$ all vanish by symmetry so only the $z$-components contribute to $T_{rr}\approx(E_z^2+B_z^2)/2$ (the sign difference relative to $T_{zz}$ is important here). Along the equator, slightly outside the bubble, both $E_z$ and $B_z$ are screened. On the other hand, just inside, only $E_z$ is screened whereas $B_z$ is enhanced (in fact, $E_z^2+B_z^2$ inside is the same as the asymptotic exterior upto and including $\mathcal{O}(\alpha)$ corrections).  Using these to compute $T_{rr}$ shows that the pressure inside is a larger positive number than that outside, implying that there is an outward force along the equator. We may therefore understand the bubble evolution by considering pressure gradients associated with these semi-classical solutions.


\subsection{Bubble expansion}\label{sec:expansion}
\begin{figure}[h]
\centering\includegraphics[width=1.0\textwidth]{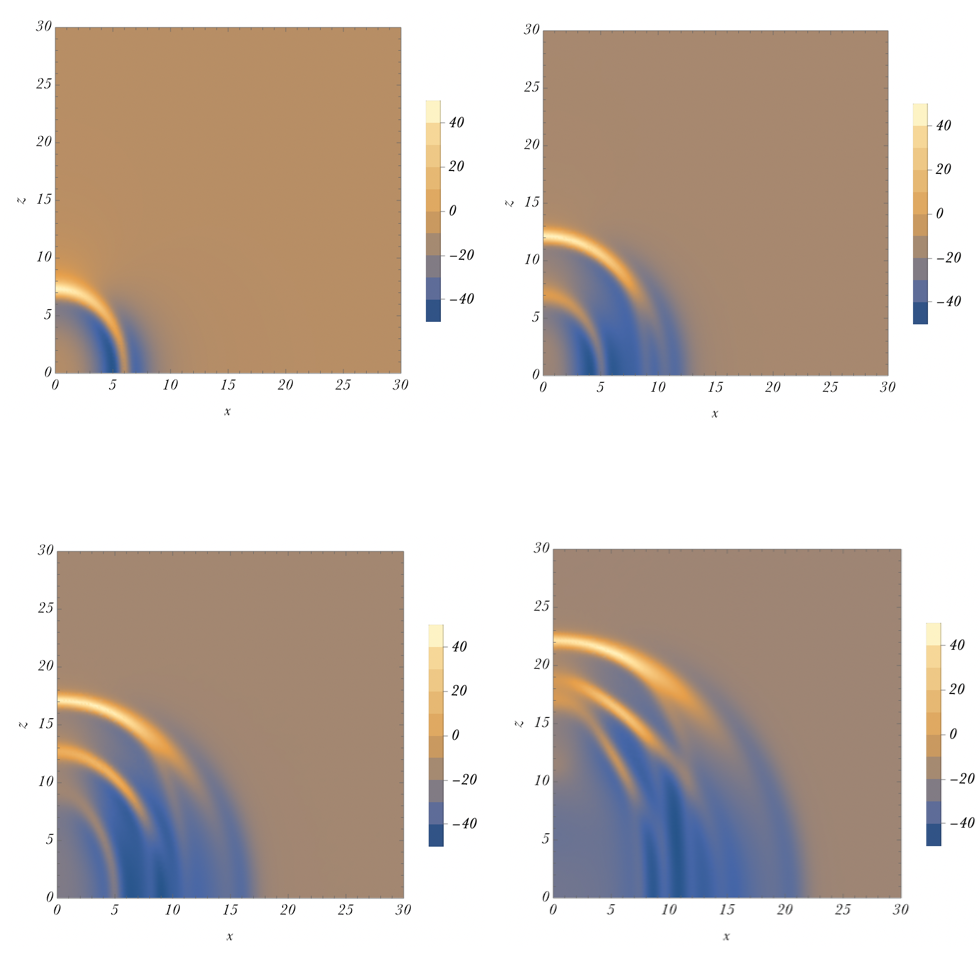}
\caption{Real time evolution of the bubble energy density (in units of $m^4$, with background subtracted) for the thick wall bubble. The top left (right) is at $t=20\ (40)$; the bottom left (right) is at $t=60 \ (80)$, in units of $m^{-1}$. The initial energy density at $t=0$ is given by figure \ref{fig:thickwallcombined}. Notice in this figure that the lower energy (dark blue) region is growing in time as the bubble expands.}
\label{fig:realtime}
\end{figure}

As a concrete example, let us study the real time evolution of the thick wall bubble of section \ref{sec:thickwall4d}. On the initial time slice, we see the nucleated oblong bubble (right hand side of figure~\ref{fig:thickwallcombined}). The bubble wall interpolates over the potential barrier and the axion field inside the bubble lies on the other side of this potential potential barrier. Since this is a thick wall profile, the axion field tunnels from the false vacuum to some intermediate value between the false vacuum and true vacuum. Subsequently, in real time, the field rolls to the true vacuum. In fact, from figure \ref{fig:realtime}, we see the axion traverses its field range three times before stabilising. In short, there is an initial quantum tunnelling, followed by two additional cascades from classical rolling. These give the domain wall an interested foliated structure that could lead to distinct observational signatures. We will study the phenomenology of these cascading decays in a more general setting in future work.

Another feature of the back-reacted time evolution is the (anti-)screening of the electromagnetic fields. In the initial configuration, the fields have values that are different from the asymptotic background although not by much. By the time the bubble has expanded, the majority of the electric field inside the bubble has been screened. The magnetic field inside the bubble is larger than its initial value but in return it is screened in a larger volume around the bubble. By integrating the energy density inside and around the bubble, we can clearly see the nucleation and evolution of the axion domain wall leads to a lower energy configuration. This is yet another piece of evidence supporting our claim that parallel $\mathbf{E}$ and $\mathbf{B}$ fields will spontaneously nucleate axion domain walls as this process leads to a lower energy state. A striking feature of the axion domain wall profile is its oblong shape. In fact, we see that this asymmetry becomes more pronounced as the domain wall evolves in time.

\section{Conclusion and outlook}
\label{sec:Conclusions}

We described a new non-perturbative process in axion electrodynamics with the axion coupled to the electromagnetic fields by means of a Chern-Simons term. This process allows parallel electric and magnetic fields to discharge by nucleating bubbles of axion domain walls that screen the fields in a complicated manner. We also described the time evolution of the axion domain wall after nucleation showing that it expands along the direction of the parallel background fields leading to elongated bubbles. The entire process is analogous to the Schwinger effect in quantum electrodynamics. Our result can be generalised in various ways. As mentioned previously, we expect that there are analogous processes in non-Abelian and higher form gauge theories. In addition, interesting effects could arise if we were to consider the process in curved spacetime or at non-zero temperature (analogous to the investigation in~\cite{Frob:2014zka} for the Schwinger effect). Returning to axion electrodynamics, and given the generic nature of the theory we are studying (for example similar theories are ubiquitous in the string Landscape~\cite{Arvanitaki:2009fg}), we expect that this process will have applications in phenomenology and formal developments alike that we now briefly discuss. 

On the phenomenological frontier, such axion domain walls could have been nucleated and annihilated in the early universe. This is similar to bubble nucleation in a first order phase transition and will produce gravitational wave signals (see for example~\cite{Croon:2023zay} and references therein) that could be within the reach of near-future gravitational wave detectors such as PTA's (such as IPTA~\cite{Antoniadis:2022pcn}), laser interferometers (such as LIGO/VIRGO~\cite{LIGOScientific:2016emj,Accadia:2011zzc}, aLIGO~\cite{LIGOScientific:2014pky} or LISA\cite{LISA:2017pwj}) and atom interferometers (such as AION~\cite{Badurina:2019hst} and MAGIS~\cite{Graham:2017pmn,MAGIS-100:2021etm}). However, there are several important features that distinguish our lopsided bubbles from those seen in typical phase transitions that are the subject of work in preparation. First, since the bubbles are elongated along the direction of the parallel electric and magnetic fields, they have less symmetry. This can allow the evasion of certain theorems that apply to spherical bubbles and prohibit the production of gravitational wave signals~\cite{Wu:1983zlp}. In summary, if one considers the expansion of two nearby spherical bubbles, the $O(2,1)$ symmetry of the system forbids the generation of gravitational waves by an analog of Birkhoff's theorem that applies to $O(3)$ symmetric spacetimes. For our non-spherical bubbles, there is generically no such $O(2,1)$ symmetry and their expansion can evade this theorem. Moreover, the possibility that the axion can traverse its field range multiple times can lead to an interesting layered structure of the axion domain wall (as pointed out in section~\ref{sec:expansion} and visible in figure~\ref{fig:realtime}). Such structures will influence the gravitational wave signals and can produce sharp peaks that are distinct from the usual broad peaks characteristic of first order phase transitions. The energy density in electromagnetic fields in and around the bubbles also displays an intricate structure that could give additional observational signatures distinct from those seen in typical first order phase transitions. These features will be the subject of work in preparation. Finally, the collision of such bubbles could lead to the formation of primordial black holes~\cite{Hawking:1982ga} and it would be interesting to see if the asymmetry can play a role in enhancing or quenching the black hole formation rate.

On the more formal side, this new discharge mechanism can be used to argue for the absence of certain objects in de Sitter space using a Festina Lente-like argument. Briefly, it was noted in \cite{Montero:2019ekk,Montero:2021otb} that light charged particles can discharge charged Nariai black holes in de Sitter space too quickly leading to a pathological Big Crunch spacetime. In a companion paper, we study a similar bound that can be derived from our new discharge process showing that it can restrict brane couplings and tensions in de Sitter space. Modulo some potential caveats, this bound will also apply to any stringy compactifications to de Sitter space. On a different frontier, there can be interesting relations between our axion domain wall solutions and the generalized symmetries literature. This is most transparent if we view the domain wall as a dynamical version of the symmetry operator. This could lead to the discovery of degrees of freedom on the domain wall that could modify decay rates or have other interesting effects if the domain wall were to interact with magnetic monopoles for instance.

\section*{Acknowledgements}
SH is grateful for the support of the Prime Minister Fellowship, Prime Minister's Office, Government of Bangladesh, as well as G-Research, London. The work of GO is supported by a Leverhulme Trust International Professorship grant number LIP-202-014. For the purpose of Open Access, the author has applied a CC BY public copyright licence to any Author Accepted Manuscript version arising from this submission.  We also also thank the STFC for support.

\appendix

\section{Bubble nucleation in $(1+1)d$ axion electrodynamics}\label{sec:2dBubbles}

In this appendix we consider false vacuum decay \cite{Coleman:1977py, Callan:1977pt, Coleman:1985rnk, brown2018schwinger, Brown:1988kg, Ai:2020vhx} in (1+1)$d$ axion electrodynamics. In this low dimensional problem, axion electrodynamics vacuum decay is very similar to Schwinger pair production and we will emphasize these similarities in this section. Much of the material in this section is analogous to the (3+1)$d$ process considered in the main body, although there are significant differences, most notably that the critical bubble and subsequent real-time evolution are trivially not lopsided.

The Lorentzian signature action for $(1+1)d$ axion electrodynamics is 
\begin{equation}
    \label{eq:amaction2d}
    I=\int d^2 x \Bigg( -\frac{1}{4}F_{\mu \nu}F^{\mu \nu} - \frac{1}{2}\partial_{\mu}\theta \partial^{\mu}\theta -\Big(\frac{m}{N}\Big)^2 \Big(1-\cos(N \theta)\Big) +  \frac{Ke}{2\pi} \theta \epsilon_{\mu \nu } F^{\mu \nu}\Bigg),
\end{equation}
where once again the integers $N$ and $K$ are the domain wall number and anomaly coefficient, respectively. 
In (1+1)$d$, there is only one independent component of the Maxwell tensor which is the electric field $E = F_{10}$. Varying the action~\eqref{eq:amaction2d} with respect to $A_\mu$ and $\theta$, we obtain the equations of motion for the fields
\begin{equation}\label{eq:eom2d}
    \partial_x E =-\frac{Ke}{\pi}\partial_x \theta, \quad \partial_t E =-\frac{Ke}{\pi}\partial_t \theta, \quad \mathrm{and,} \quad \frac{1}{m^2}\partial^2 \theta =\frac{1}{N}\Bigg(\sin(N\theta)-KN\frac{eE}{\pi m^2}\Bigg).
\end{equation}
From eq.(\ref{eq:eom2d}) constant $\theta$ local minima are given by 
\begin{equation}
\theta=N^{-1}\arcsin{(KNeE_0/\pi m^2)} 
\end{equation}
where $KNeE_0/\pi m^2 \leq 1$, with subsequent minima $2\pi/N$ units apart. We now discuss tunneling between these minima.

We would instead like to consider how charged domain walls pairs nucleate leading to electric field screening, akin to the Schwinger effect. 
Consider the theory of eq.~\eqref{eq:amaction2d} with a constant background electric field $E_0$. Using eq.~\eqref{eq:eom2d}, we see that the effective electric charge density is proportional to the spatial derivative of the axion field value and vanishes when the axion is constant. It is then easy to see that a pair of domain walls, of opposite orientation 
can screen the background electric field. This configuration can be nucleated from the vacuum in the presence of an electric field and is exactly analogous to the Schwinger pair production process with the domain walls playing the role of the electron-positron pair. The screening effect of these domain walls facilitates the decay of the background electric field. 
We describe this process in the thin and thick wall limits.

\subsection{Thin wall}

Let us start by solving the system above (i.e. eq.~\eqref{eq:eom2d} in the presence of a background electric field $E_0$) in a convenient limit where we take $K e\ll E_0$. In this limit\footnote{More precisely, we take $E \rightarrow \infty$ and $e\rightarrow 0$ keeping their product fixed. This ensures that the right hand side of the first two of eq.~\eqref{eq:eom2d} vanishes while the term proportional to $eE$ in the last of eq.~\eqref{eq:eom2d} remains relevant.} we can neglect the back-reaction of the axion profile on the electric field. This means that we can treat $E = E_0$ as a constant that solves the first two equations of~\eqref{eq:eom2d}. From the last of eq.~\eqref{eq:eom2d}, we see that the axion field evolves in the potential
\begin{equation} \label{V2d}
    V(\theta)/m^2= \frac{1}{N^2}\Big(1-\cos(N\theta)\Big)-K\frac{eE_0}{\pi m^2} \theta ~.
\end{equation}

So far, we have not assumed any thin wall limit. If we also take $KNeE_0/m^2\ll 1$, the subsequent minima will not be of significantly different energy, and the barrier height will be relatively large.
In such a scenario, the thin wall approximation is applicable and can be used to describe the tunneling process and calculate its rate~\cite{Coleman:1977py}. 

Solving $dV/d\theta=0$ one can find the minima. The first minimum is at 
\begin{align}
    \theta_0\approx \frac{KeE_0}{\pi m^2},
\end{align}
with a subsequent lower minimum at $\theta_0+2\pi/N$. Physically, when a transition occurs, a bubble forms in such a way that
in the bubble interior, $\theta=\theta_0+2\pi/N$ and in the bubble exterior $\theta=\theta_0$.\footnote{Note that this is true in the thin wall limit; when one considers a thick wall bubble, the bubble interior will be at an intermediate value of $\theta$ rather than at a minimum, after which the field classically rolls to the minimum.} Then the energy density in the interior is lower than in the exterior. A domain wall pair - with opposite orientations of the two walls - constitutes the boundary of the bubble and interpolates between these two minima. 
Because of the opposite orientations of the two walls eq.~\eqref{eq:eom2d} indicates that they carry opposite charges as expected. Moreover, this is a screening orientation: the right wall is positively charged and the left wall is negatively charged.

Let us now calculate the bubble structure and the decay rate. Under a Wick rotation $it=\tau$, the Euclidean action corresponding to eq.~\eqref{eq:amaction2d} is
\begin{equation}\label{euclideanaction2d}
    I_E=\int dx d\tau \Bigg( \frac{1}{4}F_{\mu \nu}^{(E)}F_{\mu \nu}^{(E)} +\frac{1}{2}(\partial_{\mu} \theta)^2 +\Big(\frac{m}{N}\Big)^2 \Big(1-\cos(N\theta)\Big)-i\frac{Ke}{2\pi}\theta\epsilon^{\mu \nu}F_{\mu \nu}^{(E)} \Bigg),
\end{equation}
where the superscript denotes Euclidean fields $F^{(E)}_{\mu\nu}=\pd_\mu A^{(E)}_\nu - \pd_\nu A^{(E)}_\mu$, with $iA^{(E)}_{\tau}=A_{0}$; also note that $\epsilon_{1 \tau}=1$ whereas previously in Lorentzian signature $\epsilon_{01}=1$.
The equations of motion, upon varying with respect to $A^{(E)}$ and $\theta$, are
\begin{equation}\label{euclideaneom2d}
    \partial_\mu F^{(E)\mu\nu}=i\frac{Ke}{\pi}\epsilon^{\mu \nu}\partial_\mu \theta, \quad \mathrm{and}\quad \partial^2 \theta =m^2 \Bigg(\frac{\sin{(N\theta)}}{N}-i\frac{Ke}{2\pi m^2}\epsilon^{\mu \nu}F^{(E)}_{\mu\nu}\Bigg).
\end{equation}
First we search for an instanton when $KNeE_0/m^2\ll 1$, and $Ke/N \ll E_0$. Then the screening effect of the electric field by the domain wall is small. In essence, we take the right hand side of the first of eq.~\eqref{euclideaneom2d} to be vanishingly small, so that the electric field is effectively constant. Then the photon just serves as a background field. Analogous to the Schwinger effect, we attempt an $O(2)$ symmetric ansatz, taking $\theta=\theta(\rho)$ where we define the dimensionless length $\rho=m\sqrt{\tau^2+x^2}$. Then the equation of motion for the axion becomes 
\begin{equation}\label{eq:axion2dthinwalleom}
    \frac{d^2 \delta\theta}{d \rho^2}+\frac{1}{\rho}\frac{d\delta\theta}{d\rho}=\frac{1}{N}\sin(N\delta\theta)+\mathcal{O}(\theta_0),
\end{equation}
\\
where $\delta\theta(\rho)=\theta(\rho)-\theta_0$. In the thin wall limit where $\theta_0 \ll 1$ we may drop the $\mathcal{O}(\theta_0)$ term, as well as the friction term on the left hand side, which means~\eqref{eq:axion2dthinwalleom} is solved by the sine-Gordon soliton
\begin{equation}
    \theta(\rho)=\theta_0+\frac{2\pi}{N}-\frac{4}{N}\arctan{(\exp{(\rho-\rho_*)}}.
\end{equation}
The wall thickness is therefore $\mathcal{O}(m^{-1})$, and the wall is situated at $x=m^{-1} \rho_*$ upon nucleation. We may find $\rho_*$ by substituting this solution in eq.~\eqref{euclideanaction2d}, and minimizing\footnote{Recall that we are keeping the electric field constant in this approximation. We will drop this condition in an example momentarily.} to obtain the radius of the critical bubble. We find 
\begin{align}
\rho_* \approx \frac{4m^2}{NKeE_0}. 
\end{align}
This expression is in good agreement with numerical solutions in the regime of validity of the thin wall approximation.
Moreover, a solution with this critical radius has Euclidean action
\begin{align}
    I_E= \frac{32\pi m^2}{N^3 K e E_0}
\end{align}
Already, we see a resemblance with the Schwinger tunnelling exponent $I_{\rm Schwinger} = \pi \mu^2/qE$. To understand this point better, it is helpful to know the energy momentum tensor in order to compute the domain wall mass. The Hilbert energy momentum tensor is 
\begin{equation}\label{eq:EMtensor2d}
T^{\mu\nu}=F^{\mu\lambda}F^{\nu}_{\ \lambda}+\partial^{\mu}\theta \partial^{\nu}\theta-\eta^{\mu \nu}\left(\frac{1}{4}F_{\alpha\beta}F^{\alpha\beta}+\frac{1}{2}(\partial\theta)^2 +\left(\frac{m}{N}\right)^2 \Big(1-\cos{(N\theta)}\Big)\right),
\end{equation}
which agrees with the canonical energy-momentum tensor upon Belinfante improvement; note the absence of the topological term\footnote{Actually even under Belinfante improvement, the canonical energy momentum tensor will appear to have terms arising from the topological Chern-Simons operator in the action, but one can show that such terms actually cancel, and we are left with the Hilbert form.}. It is tempting to infer from this observation that the Chern-Simons term is immaterial, but such a conclusion is misleading: the equations of motion, and their solutions $\theta$ and $A_\mu$, depend on the topological term and so ultimately the energy-momentum tensor does too when evaluated on a solution. Using eq.~\eqref{eq:EMtensor2d}, we find that the mass of the domain wall is $m_{DW}\approx 8m/N^2$. Here again, the thin wall approximation is necessary. The domain wall charges are $Q_{DW}=\pm 2eK/N$ from integrating the first of eq.~\eqref{eq:eom2d}. It then follows that the decay rate is 
\begin{equation}
    \Gamma/V\sim \exp{\left(-\frac{\pi m_{DW}^2}{Q_{DW}E_0}\right)}.
\end{equation}
The exponent of the nucleation rate is similar to that in the Schwinger effect of charged particles, except with the particle's mass and charge replaced by those of the the domain wall. This similar scaling behavior demonstrates that axion domain wall nucleation is a type of Schwinger process. 

There are a few points that we wish to emphasize here. First, in the case of the Schwinger effect, one nucleates pairs of charged particles. Here, one instead nucleates from the vacuum a pair of extended objects - domain walls - which have an effective electric charge. In (1+1)$d$ the domain wall is a (0+1)$d$ object and this process is thus very similar to particle production in the Schwinger effect. In higher dimensions this conceptual difference becomes more important as the domain walls can no longer be thought of as particles. Importantly, this process can happen even in the absence of charged particles in the spectrum. 

Moreover, as we stated earlier, the magnitude of the charge on the wall is $2eK/N$, yet we have neglected the change in the electric field. However, in practice the interior electric field is now $E=E_0-eK/N$, and after the walls have propagated outwards, the new background electric field is $E_0-eK/N$, and the expectation value of the axion is $\theta=N^{-1}\arcsin{(KNe(E_0-eK/N)/m^2)}$. We only considered cases for which $E_0\gg eK/N$ so that the screening of the electric field is small for each individual bubble that is nucleated. This approximation is not essential. Nevertheless it leads to an important matter regarding bubble expansion in real time after nucleation. 

We may understand the outward expansion as being caused by the pressure difference between the interior and the exterior. In particular, from the energy momentum tensor in eq.~\eqref{eq:EMtensor2d} the pressure in a region where the axion is not varying significantly is 
\begin{align}
P\approx -\frac{E^2}{2}-\Big(\frac{m}{N}\Big)^2\Big(1-\cos{(N\theta)}\Big). 
\end{align}
This applies in the interior and exterior regions, but not on the wall itself. This expression need not hold in the thick wall limit in which the interior axion field is not constant. As $\theta$ changes by $2\pi/N$ between the exterior and the interior, the second term in the expression for the pressure $P$ does not change. However, the electric field is screened on the inside and the pressure is therefore less negative (i.e. higher) inside the bubble. This pressure difference leads to an outward force on the wall which causes expansion, eventually leading to a hyperbolic trajectory in real time.

\subsection{Thick wall}

Now, we may study processes in which the thin wall approximation is no longer valid. In this case, we consider the scenario in which the difference in energies of the two minima is no longer small compared to the barrier height. To add to this, we do not assume that the electric field is unchanged, and we incorporate the screening effect of the domain wall. Then the axion will begin rolling in the inverted potential subject to the (Euclidean) time dependent friction. This will lead to a thick wall profile for which one must numerically integrate the equations of motion to obtain the instanton solution. The bounce action is then given by substituting this solution in the Euclidean action. 

There is subtle point that we should clarify before proceeding. This point stems from the fact that for a dynamical electric field we have to treat the action, in particular the boundary terms, carefully~\cite{Brown:1988kg}. Typically, when we vary the action with respect to the fields, it is assumed that the field configurations are fixed on the boundary. This allows us to drop boundary terms. Then requiring a stationary action immediately yields the equations of motion. In our case, we would like the fields $\theta$ and $F_{\mu\nu}$ (or $E$) to be fixed at the boundary. The standard procedure however assumes that $A_\mu$ is fixed at the boundary rather than $F_{\mu\nu}$. We can see this by varying eq.~\eqref{eq:amaction2d} with respect to $A_\mu$; note that the boundary term contains $\delta A_\mu$ which one typically assumes vanishes\footnote{Actually, the component of $A_{\mu}$ tangential to the boundary is required to be fixed on the boundary.}. We may rectify this issue by adding a total derivative $\partial_\mu ((F^{\mu \nu}-Ke\theta\epsilon^{\mu \nu}/\pi)A_\nu)$ to the Lagrangian of eq.~\eqref{eq:amaction2d}\footnote{This addition amounts to a Legendre transformation at the level of $\delta I$, since $Ke\theta\epsilon^{\mu \nu}/\pi-F^{\mu \nu}$ is the momentum conjugate to $A_\nu$.}. Varying the new action, one can show that the surface term only contains $\delta F_{\mu\nu}$ now, and not $\delta A_\mu$. This procedure does not alter the equations of motion since we are merely adding a total derivative. Therefore, solutions, in particular bubble configurations, remain unchanged. Nevertheless, quantities that depend not only on the functional form of the action but also its numerical value - such as decay rates - may be altered. Explicitly, the new action $I'$ is 
\begin{align}\label{action2dwithLegendre}
    I' &=\int d^2x\left(\frac{1}{4}F_{\mu \nu}F^{\mu \nu}-\frac{1}{2}(\partial\theta)^2-\left(\frac{m}{N}\right)^2(1-\cos{(N\theta)})\right)\\
    &\ \ \ \qquad +\int d^2x \left(A_\nu\partial_\mu \left(F^{\mu\nu}-\frac{Ke}{\pi}\theta\epsilon^{\mu\nu}\right)\right).\nonumber
\end{align}
It is a simple exercise to show that varying this action with respect to the fields gives the correct equations of motion, with the requirement that $\delta \theta$ and $\delta F_{\mu\nu}$ vanish on the boundary. One can Wick rotate eq. (\ref{action2dwithLegendre}) to compute the decay rate. The second line vanishes on the equations of motion. Note that the topological term no longer explicitly appears in the first line -- which is the line that contributes to the decay rate -- despite playing a crucial role in the decay process. Let us now return to the problem of computing bubble nucleation rates for arbitrary walls.

\begin{figure}[h]
\centering\includegraphics[width=0.6\textwidth]{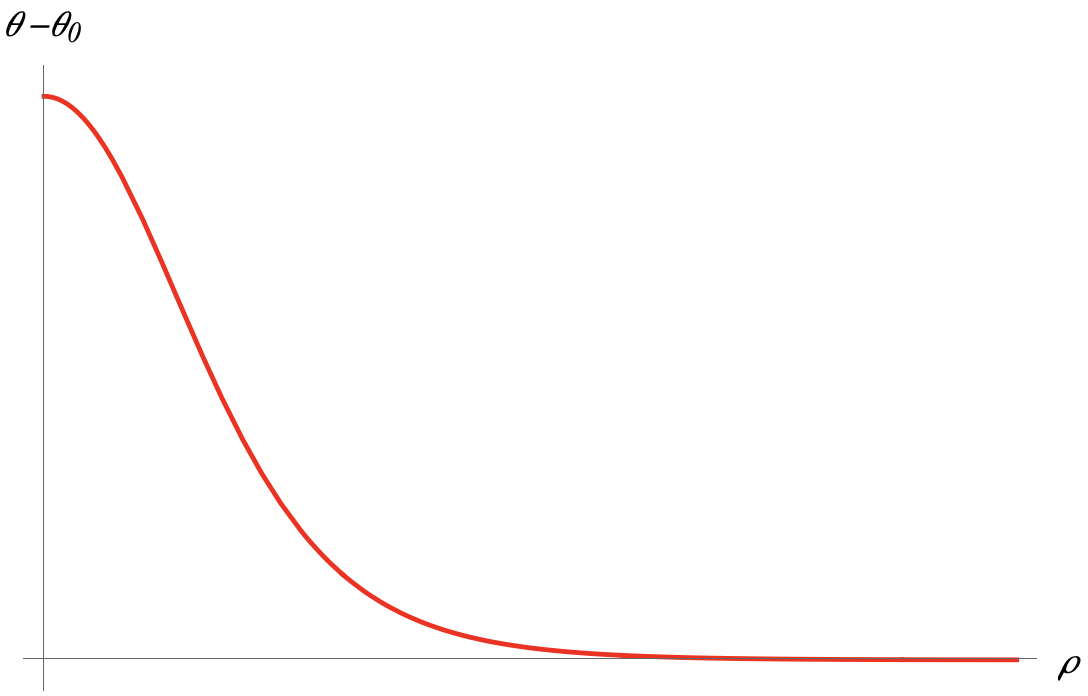}
\caption{Bubble profile for $\theta-\theta_0$.}
\label{fig:deltatheta2d}
\end{figure}

First we will assume an $O(2)$ symmetric ansatz for our fields, so that $\theta=\theta(\rho)$ as before, and $A=A_\rho(\rho) d\rho +A_\phi(\rho) d\phi.$ where $\phi$ is the angular coordinate in the $\tau x$-plane. Moreover, we gauge fix\footnote{This gauge choice is consistent with the $O(2)$ ansatz under which the gauge fields and transformation parameters do not depend on $\phi$ either. For instance, we may not have been able to consistently require $A_\phi$ to vanish in general by gauge choice.} $A_\rho=0$. Then it automatically follows that $E=E(\rho)$ since now $E= i\rho^{-1}\partial_\rho A_\phi(\rho)$. We will apply a similar strategy to dynamics in $(3+1)d$ in section \ref{sec:O2O2solution}. Varying eq.~\eqref{euclideanaction2d}, subject to the boundary condition at infinity, $\theta_0=N^{-1}\sin^{-1}(NKeE_0 /\pi m^2)$, we find the following equation for $\delta \theta$:
\begin{equation}\label{axioneom2d}
    \frac{d^2 \delta \theta}{d\rho^2}+\frac{1}{\rho}\frac{d \delta \theta}{d\rho}-\frac{1}{N}\left(\sin(N\theta_0+N\delta \theta)-\sin(N \theta_0)\right)-\left(\frac{Ke}{\pi m}\right)^2 \delta \theta=0,
\end{equation}
where we must also require that the derivative of $\delta \theta$ vanishes at the origin. Note that we have integrated the first of eq. (\ref{euclideaneom2d}) and substituted the result into the second. Then numerical integration gives the axion profile. As an example, consider the case in which $E_0/m=3$, and $e/m= N=K=1$. Numerical integration of eq. (\ref{axioneom2d}) shows that at the center of the bubble, $\theta-\theta_0 \approx  1.955$ and  figure \ref{fig:deltatheta2d} explicitly shows the thick wall bubble profile.
Substituting this profile in eq. (\ref{action2dwithLegendre}) gives the decay rate $\Gamma/V\sim\exp(-B)$, with $B \approx5.416$. This is a reasonably fast decay process since the exponent is approaching unity. Note though, that when $B\lesssim 1$, we do not trust the semi-classical analysis but intuitively the decay is also fast for $B\lesssim 1$. An important aspect of this decay process is that even when we included the back-reaction of the electric field, the $O(2)$ ansatz was still valid. In section \ref{sec:4dBubbles} we will find that in $(3+1)d$, incorporating the back-reaction of the gauge field drastically reduces the symmetry group of the solution.

\section{Back-reaction equations}\label{sec:back-reaction}

Let us return to the problem of solving the Euclidean equations of motion~\eqref{eq:Euclideaneomaxion} and~\eqref{eq:Euclideaneomphoton} which are reproduced below (with $A^\mu\rightarrow \sqrt{\alpha}A^\mu/m$): 
\begin{align}
    \partial_\rho^2 \theta+\frac{\partial_\rho \theta}{\rho}+ \partial_r^2 \theta+\frac{\partial_r \theta}{r}-\frac{\sin(N\theta)}{N}+\frac{i K}{\pi r\rho}\left(\frac{m}{f}\right)^2 \left(\partial_\rho A_\sigma \partial_r A_\phi - \partial_r A_\sigma \partial_\rho A_\phi \right)&=0, \label{eq:eomo2o21B}\\
    \partial_\rho^2 A_\sigma-\frac{\partial_\rho A_\sigma}{\rho}+ \partial_r^2 A_\sigma+\frac{\partial_r A_\sigma}{r}-\frac{iK\alpha}{\pi}\frac{\rho}{r} \left(\partial_r \theta \partial_\rho A_\phi - \partial_\rho \theta \partial_r A_\phi \right)&=0, \label{eq:eomo2o22B}\\
    \partial_\rho^2 A_\phi+\frac{\partial_\rho A_\phi}{\rho}+ \partial_r^2 A_\phi-\frac{\partial_r A_\phi}{r}-\frac{iK\alpha}{\pi}\frac{r}{\rho} \left(\partial_r \theta \partial_\rho A_\sigma - \partial_\rho \theta \partial_r A_\sigma \right)&=0. \label{eq:eomo2o23B}
\end{align}
In order to do that, we have to impose appropriate boundary conditions on the fields. Far away from the origin (i.e. when $\rho \rightarrow \infty$ or $r \rightarrow \infty$), all fields must match on to their asymptotic values. These are the ones given in eqs.~\eqref{eq:EfieldBC} for $A_\phi$ and $A_\sigma$. For the axion field $\theta$, it should match to a value $\theta_0$ given by a solution to eq.~\eqref{eq:potminima}. In addition, all fields must obey vanishing Neumann boundary conditions at $r = 0$ or $\rho=0$ to avoid spurious singularities. Finally, we also impose the Dirichlet conditions $A_\phi = 0$  at $\rho=0$ and $A_\sigma = 0$ at $r = 0$ since these determine components of the $\mathbf{E}$ and $\mathbf{B}$ fields that vanish by the symmetry of the problem. We show two examples of the solutions to these equations in the next subsection. Note the presence of the features discussed in section~\ref{sec:WittenSikivie}, i.e. the elongation of the instanton solution and the (anti-)screening of the electric and magnetic fields. 

\subsection{Gauge field}
The interaction of the axion with the gauge field will induce changes in both these fields. Concretely, we denote these changes through the following perturbations: $\theta(\rho,r)=\tilde{\theta}(u)+\alpha\delta \theta (\rho,r), A_{\phi}(\rho,r)=\tilde{A}_{\phi}(\rho)+\alpha\delta A_{\phi }(\rho,r)$, and $ A_{\sigma }(\rho,r)=\tilde{A}_{\sigma}(r)+\alpha\delta A_{\sigma}(\rho,r)$, with the expansion truncated to first order in $\alpha$. 

It then follows from eq.~(\ref{eq:eomo2o22B}) and eq.~(\ref{eq:eomo2o23B}) that the gauge field perturbations satisfy the following differential equations at $\mathcal{O}(\alpha)$:
\begin{align}
    \partial_\rho^2 \delta A_\sigma-\frac{\partial_\rho \delta A_\sigma}{\rho}+ \partial_r^2 \delta A_\sigma+\frac{\partial_r \delta A_\sigma}{r}&=-\frac{iK}{\pi}\frac{\tilde{B_0}}{m^2}\frac{\rho^2}{\sqrt{\rho^2 +r^2}}  \frac{d\tilde{\theta}}{du},\label{Apert1} \\
    \partial_\rho^2 \delta A_\phi+\frac{\partial_\rho \delta A_\phi}{\rho}+ \partial_r^2 \delta A_\phi-\frac{\partial_r \delta A_\phi}{r}&=\frac{K}{\pi}\frac{\tilde{E_0}}{m^2}\frac{r^2}{\sqrt{\rho^2 +r^2}}  \frac{d\tilde{\theta}}{du}.\label{Apert2} 
\end{align}
Note that the terms on the right hand sides of both equations above show that the varying axion profile sources the change in the electric and magnetic fields from their unperturbed background values. We also find that at corrections linear in $\alpha$, the components of the gauge field perturbation effectively decouple. (In general, gauge field components will not decouple order by order with higher powers of $\alpha$.) We must now numerically integrate these equations, subject to the requirements above.
\subsection{Axion}
We may also compute the change in the bubble profile due to these back-reactions of the electromagnetic field. From eq.~\eqref{eq:eomo2o21B}, at $\mathcal{O}(\alpha)$,
\begin{equation}\label{eq:deltathetaB}
    \partial_\rho^2 \delta\theta+\frac{\partial_\rho \delta\theta}{\rho}+ \partial_r^2 \delta\theta+\frac{\partial_r \delta\theta}{r}=\delta\theta\cos\tilde\theta-\frac{1}{\pi}\Big(\frac{m}{f}\Big)^2\Bigg(\frac{\tilde{E}_0}{m^2}\frac{\partial_r \delta A_\phi}{r}+\frac{\tilde{B}_0}{m^2}\frac{\partial_\rho \delta A_\sigma}{\rho}\Bigg),
\end{equation}
which upon integrating subject to the same Neumann boundary conditions as above, we obtain a modified bubble profile, which incorporates $O(2)\times O(2)$ corrections to the leading order $O(4)$ solution.

\bibliographystyle{JHEP}
\bibliography{refs.bib}

\end{document}